\documentclass[twocolumn,showpacs,showkeys,preprintnumbers,amsmath,amssymb]{revtex4}

\usepackage{graphicx}
\usepackage{dcolumn}
\usepackage{bm}

\newcommand{\ti}{\'{\i}}

\newcommand{\emp}{\begin{equation}}
\newcommand{\fin}{\end{equation}}
\newcommand{\empn}{\begin{equation*}}
\newcommand{\finn}{\end{equation*}}

\newcommand{\bea}{\begin{eqnarray}}
\newcommand{\eea}{\end{eqnarray}}
\newcommand{\eger}{\begin{gather}}
\newcommand{\fger}{\end{gather}}
\newcommand{\egn}{\begin{gather*}}
\newcommand{\fgn}{\end{gather*}}
\newcommand{\bit}{\begin{itemize}}
\newcommand{\eit}{\end{itemize}}
\newcommand{\ket}[1]{\vert{#1}\rangle}
\newcommand{\bra}[1]{\langle{#1}\vert}

\newcommand{\lrp}[1]{\left(#1\right)}
\newcommand{\lrpr}[1]{\left[#1\right]}
\newcommand{\lrpy}[1]{\left\{#1\right\}}

\newcommand{\al}{\ensuremath{{\alpha}}}

\newcommand{\be}{\ensuremath{{\beta}}}
\newcommand{\ga}{\ensuremath{{\gamma}}}

\newcommand{\la}{\ensuremath{{\lambda}}}

\newcommand{\h}{\ensuremath{{\chi}}}

\newcommand{\0}{\phantom{0}}
\newcommand{\m}{\llap{$-$}}

\newcommand{\depar}[3]{\ensuremath{\frac{\partial^{#1}{#2}}{\partial{#3}^{#1}}}}

\newcommand{\JPA}{{ J. Phys. A: Math. Gen.} }
\newcommand{\NP}{{ Nucl. Phys.} }
\newcommand{\PL}{{ Phys. Lett.} }

\newcommand{\RMP}{{ Rev. Mod. Phys.} }
\newcommand{\JMP}{{\em J. Math. Phys.} }
\newcommand{\F}{\ensuremath{\lrpr{3}_q\sqrt{\frac{\lrpr{3}_q+\lrpr{2}_q}{\lrpr{4}_q+1}}}}
\newcommand{\G}{\ensuremath{C\lrp{q}-E\lrp{q}-D\lrp{q}}}
\newcommand{\Ha}{\ensuremath{q^{-2}E\lrp{q}+\frac{q^2}{2}\lrp{A\lrp{q}+\frac{\lrp{\lrpr{\frac{3}{2}}_q}^2\sqrt{\lrpr{5}_q}}{\lrpr{3}_q\lrpr{\frac{1}{2}}_q\lrpr{\frac{5}{2}}_q}B\lrp{q}}}}
\newcommand{\I}{\ensuremath{q^{2}E\lrp{q}+\frac{q^3}{2}\lrp{A\lrp{q}-\frac{\lrp{\lrpr{\frac{3}{2}}_q}^2\sqrt{\lrpr{5}_q}}{\lrpr{3}_q\lrpr{\frac{1}{2}}_q\lrpr{\frac{5}{2}}_q}B\lrp{q}}}}

\begin{document}


\title{The Gell-Mann-Okubo and Coleman-Glashow relations for octet and decuplet baryons in the $SU_q(3)$ quantum algebra}

\author{Antonio E C\'arcamo Hern\'andez}
\email{aecarcamoh@unal.edu.co}
\affiliation{%
Departamento de F{\ti}sica, Universidad Nacional de Colombia, Bogot\'a, Colombia} 
\altaffiliation{Present address: Cra 30 No. 45-03, Ed 405, Ciudad Universitaria, Departamento de F{\ti}sica, Universidad Nacional de Colombia, Bogot\'a, Colombia. Phone: 057-1-3165000 ext 13080-13082.}
%


\date{\today}

\begin{abstract}
The $q$-deformed Clebsch-Gordan coefficients corresponding to the $\lrpy{3}\times\lrpy{21}$ reduction of the $SU_q(3)$ quantum algebra are computed. From these results and using the quantum Clebsch-Gordan coefficients for the $\lrpy{21}\times\lrpy{21}$ reduction found by Z.Q.Ma, the $q$-deformed Gell-Mann-Okubo mass relations for octet and decuplet baryons are determined by generalizing the procedure used for the $SU(3)$ algebra. We also determine the Coleman-Glashow relations for octet and decuplet baryons in the $SU_q(3)$ algebra. Finally, by using the experimental particle masses of the octet and decuplet baryons, two values of the $q$-parameter are found and adjusted for the predicted expressions of the masses (one for the Gell-Mann-Okubo mass relations and the other for the Coleman-Glashow relations) and a possible physical interpretation is given.
\end{abstract}

\maketitle

\section{\label{sec:level1}Introduction}
The $q$-deformed algebras extend the domain of classical group theory and constitute a new and growing field of mathematics with vast potential for applications in physics \cite{Drinfeld,Reshetikhin,Biedenharn,Jaganathan,Macfarlane,Delgado}. 
The quantum algebra $SU_q(3)$, as a generalization of the $SU(3)$ algebra, has been studied by several authors \cite{Ma,Rodriguez,Quesne,Yu,Carcamo} with several applications in particle physics, conformal field theory, statistical mechanics, quantum optics, condensed matter, molecular, atomic and nuclear spectroscopy \cite{Biedenharn,Jaganathan,Gavrilik,Carcamo}. In these applications either an existing model is identified with a quantum algebraic structure or a standard model is deformed to show a underlying quantum algebraic structure which reveals new features \cite{Jaganathan}.\\
Applications of quantum algebras in particle physics have been explored in several works \cite{Gavrilik}. In hadronic phenomenology $q$-deformed mass relations between particle families in the octet and decuplet baryons have been determined from the computation of the mass operator's expectation value. In these works, the mass operator has been defined in terms of generators of $SU_q(4)$ and $SU_q(5)$ and its expectation value has been computed from the determination of their matrix elements.\\        
Our aim in this article is to determine the $q$-deformed Gell-Mann-Okubo and Coleman-Glashow relations for octet and decuplet baryons in the $SU_q(3)$ quantum algebra. First, we develop the $SU_q(3)$ quantum algebra and use it to compute the $q$-deformed Clebsch-Gordan coefficients corresponding to the $\lrpy{3}\times\lrpy{21}$ reduction. Second, to derive the $q$-deformed Gell-Mann-Okubo mass relations for octet and decuplet baryons in the $SU_q(3)$ quantum algebra we generalize the traditional procedure for the $SU(3)$ algebra \cite{Swart} and use the previous results together with the quantum Clebsch-Gordan coefficients corresponding to the $\lrpy{21}\times\lrpy{21}$ reduction \cite{Ma}. After that, we obtain the $q$-deformed Coleman-Glashow relations for octet and decuplet baryons by following the same procedure used for the $SU(3)$ algebra \cite{Carruthers}. Finally, by using the experimental particle masses of the octet and decuplet baryons, two values of the $q$-parameter are found and adjusted for the Gell-Mann-Okubo and Coleman-Glashow mass relations and a possible physical interpretation is given.

\section{The $SU_q(3)$ quantum algebra}
The quantum algebra $SU_q(3)$ is generated by the ope-rators  $\widehat{I}$,
$\widehat{h}_1$, $\widehat{h}_2$, $\widehat{X}^{\pm}_1$, $\widehat{X}^{\pm}_2$, which satisfy the following commutation relations \cite{Rodriguez}:
\begin{eqnarray}
\lrpr{\widehat{h}_i,\widehat{h}_j}=0,\hspace{1.2cm}\lrpr{\widehat{h}_i,\widehat{X}^{\pm}_j}=\pm a_{ij}\widehat{X}^{\pm}_j\\
\lrpr{\widehat{X}^+_i,\widehat{X}^-_j}=\delta_{ij}\lrpr{\widehat{h}_i}_q=\delta_{ij}\frac{q^{\widehat{h}_i}-q^{-\widehat{h}_i}}{q-q^{-1}}\hspace{0.25cm}i,j=1,2
\end{eqnarray}
together with: 
\emp
\lrp{\widehat{X}^{\pm}_i}^2\widehat{X}^{\pm}_j+\widehat{X}^{\pm}_j\lrp{\widehat{X}^{\pm}_i}^2-\lrpr{2}_q\widehat{X}^{\pm}_i\widehat{X}^{\pm}_j\widehat{X}^{\pm}_i=0\hspace{0.5cm}i\ne j
\fin
where $a_{ij}$ is the Cartan matrix given by:
\emp
a_{ij}=
\lrp{\begin{array}{rl}
2 & -1\\
-1 & \hspace{0.3cm}2
\end{array}}
\fin

Additional generators $\widehat{X}^{\pm}_3$ are introduced by \cite{Quesne}:
\emp
\widehat{X}^{+}_{3}=q^{1/2}\lrpr{\widehat{X}^+_1,\widehat{X}^+_2}_{q^{-1}},\hspace{0.6cm}\widehat{X}^{-}_{3}=-q^{-1/2}\lrpr{\widehat{X}^-_2,\widehat{X}^-_1}_{q}\fin
where:
\begin{eqnarray}
\widehat{X}^{\pm}_1=\widehat{T}_{\pm},\hspace{1cm}\widehat{X}^{\pm}_2=\widehat{U}_{\pm},\hspace{1cm}\widehat{h}_1=2\widehat{T}_3,\\
\widehat{h}_2=-\widehat{T}_3+\frac{3}{2}\widehat{Y},\hspace{1cm}\lrpr{\widehat{A},\widehat{B}}_q=\widehat{A}\widehat{B}-q\widehat{B}\widehat{A}
\end{eqnarray}

From the previous expressions, we obtain: 

\begin{eqnarray}
\lrpr{\widehat{X}^{+}_3,\widehat{X}^{-}_3}=-\frac{q^{\widehat{h_3}}-q^{-\widehat{h_3}}}{q-q^{-1}}=-\lrpr{\widehat{h}_3}_q,\hspace{1.95cm}\\
\lrpr{\widehat{h}_i,\widehat{X}^{\pm}_3}=\pm\widehat{X}^{\pm}_3\hspace{0.35cm}\textit{With}\hspace{0.35cm}\widehat{h}_3=\widehat{h}_1+\widehat{h}_2=\widehat{T}_3+\frac{3}{2}\widehat{Y}
\end{eqnarray}

According to the standard coproduct definition at $SU_q(3)$, the following expressions are obtained \cite{Carcamo}:

\begin{eqnarray}
\Delta\widehat{T}_{\pm}=\widehat{T}_{\pm}\otimes q^{\widehat{T}_3}+q^{-\widehat{T}_3}\otimes\widehat{T}_{\pm},\hspace{2.4cm}\\
\Delta\widehat{U}_{\pm}=\widehat{U}_{\pm}\otimes q^{\lrp{3\widehat{Y}-2\widehat{T}_3}/4}+q^{-\lrp{3\widehat{Y}-2\widehat{T}_3}/4}\otimes\widehat{U}_{\pm}
\end{eqnarray}

Hence, the coproduct of two $SU_q(3)$ irreps is defined as \cite{Carcamo}:      \begin{align}
\begin{split}
\widehat{T}_{\pm}\lrp{\psi^{\lrpy{\la_1,\la_2}}_{\eta}\psi^{\lrpy{\mu_1,\mu_2}}_{\be}}=\lrp{q^{-\widehat{T}_3}\psi^{\lrpy{\la_1,\la_2}}_{\eta}}\lrp{\widehat{T}_{\pm}\psi^{\lrpy{\mu_1,\mu_2}}_{\be}}\\+\lrp{\widehat{T}_{\pm}\psi^{\lrpy{\la_1,\la_2}}_{\eta}}\lrp{q^{\widehat{T}_3}\psi^{\lrpy{\mu_1,\mu_2}}_{\be}}\hspace{0cm}\nonumber
\end{split}
\end{align}
\begin{align}
\begin{split}
\widehat{U}_{\pm}\lrp{\psi^{\lrpy{\la_1,\la_2}}_{\eta}\psi^{\lrpy{\mu_1,\mu_2}}_{\be}}\hspace{5cm}\\
\hspace{0.4cm}=\lrp{\widehat{U}_{\pm}\psi^{\lrpy{\la_1,\la_2}}_{\eta}}\lrp{q^{\lrp{3\widehat{Y}-2\widehat{T}_3}/4}\psi^{\lrpy{\mu_1,\mu_2}}_{\be}}\\
\hspace{0.5cm}+\lrp{q^{-\lrp{3\widehat{Y}-2\widehat{T}_3}/4}\psi^{\lrpy{\la_1,\la_2}}_{\eta}}\lrp{\widehat{U}_{\pm}\psi^{\lrpy{\mu_1,\mu_2}}_{\be}}\nonumber
\end{split}
\end{align}
where $\psi^{\lrpy{\nu}}_{\al}$, with $\al=1,2,\cdots dim\lrp{\lrpy{\nu}}$, is the state with eigenvalues $t_{\al}$, $t_{3\al}$ and $y_{\al}$, which belongs to the $SU_q(3)$ representation $\lrpy{\nu}=\lrpy{\nu_1,\nu_2}$ \footnotemark[1]

\footnotetext[1]{In this article, we consider $q\in R$ following the Quesne's pres-cription.}


\section{The $q$-deformed Gell-Mann-Okubo mass relations for octet and decuplet baryons}

In order to get expressions of the masses corresponding to the octet and decuplet baryons, we introduce the following mass operator:
\emp
\widehat{M}=\widehat{M}_S+\widehat{M}_T   
\label{b1}
\fin
where $\widehat{M}_S$ is a $SU_q(3)$ invariant and $\widehat{M}_T$ an isospin escalar.
  
By computing the expectation value of the mass ope-rator in a $\ket{\psi^{\lrpy{\la}}_{\al}}$ state belonging to a $\lrpy{\la}$ representation, we obtain that the mass of the particle corresponding to this state is given by:

\emp
m\lrp{\lrpy{\la},\lrpy{\mu}}=\bra{\psi^{\lrpy{\la}}_{\al}}\widehat{M}_S\ket{\psi^{\lrpy{\la}}_{\al}}+\bra{\psi^{\lrpy{\la}}_{\al}}\widehat{M}_T\ket{\psi^{\lrpy{\la}}_{\al}}
\label{b2}
\fin
with:

\begin{eqnarray}
\ket{\psi^{\lrpy{\la}}_{\al}}&=&\ket{\lrpy{\la},\lrpy{\mu}}=\ket{\lrpy{\la},y,t,t_z}\nonumber\\
\lrpy{\mu}&=&\lrpy{y,t,t_z} 
\end{eqnarray}
where in order to simplify the notation we have taken $y=y_{\al}$, $t=t_{\al}$ and $t_z=t_{3\al}$ being $y$, $t$ and $t_z$ the hypercharge, the isospin and the $z$ isospin component respectively.\\

As $\widehat{M}_S$ is a $SU_q(3)$ scalar, its expectation value is the same for all members of the multiplet corresponding to a $\lrpy{\la}$ representation. Therefore, acccording to the previous, we obtain:

\emp
\bra{\psi^{\lrpy{\la}}_{\al}}\widehat{M}_S\ket{\psi^{\lrpy{\la}}_{\al}}=m_S\lrp{\lrpy{\la}}
\label{b3}
\fin 

We also note that the $\widehat{M}_T$ operator can be written as an expansion of irreducible tensor operators of $SU_q(3)$, according to:  

\emp
\widehat{M}_T=\sum_{\mu\nu}\widehat{T}^{\lrpy{\nu}}_{\mu}
\label{b4}
\fin

Then, by taking into account that $\lrpr{\widehat{M}_{T},\widehat{Y}}=\lrpr{\widehat{M}_{T},\widehat{T}}=0$, from the previous expression we obtain that the irreducible tensor operators $\widehat{T}^{\lrpy{\la}}_{\mu}$ should satisfy the following conditions:

\emp
\lrpr{\widehat{T}^{\lrpy{\nu}}_{\mu},\widehat{Y}}=0,\hspace{1cm}\lrpr{\widehat{T}^{\lrpy{\nu}}_{\mu},\widehat{T}}=0
\label{b5}
\fin

On the other hand, according to the Wigner-Eckart theorem for the quantum group $SU_q(3)$, we have that the matrix element of the irreducible tensor operator $\widehat{T}^{\lrpy{\la_2}}_{\mu_2}$ is given by \cite{Klimyk}:

\begin{widetext}
\emp
\bra{\lrpy{\la_3},y_3,t_3,t_{3z}}\widehat{T}^{\lrpy{\la_2}}_{\mu_2}\ket{\lrpy{\la_1},y_1,t_1,t_{1z}}=\sum_{\ga}\begin{pmatrix}
\la_1 & \la_2 & \la_{3\ga}\\
\mu_1 & \mu_2 & \mu_3
\end{pmatrix}_q\bra{\lrpy{\la_3}}|\widehat{T}^{\lrpy{\la_2}}|\ket{\lrpy{\la_1}}_{\ga}
\label{b6}
\fin
\end{widetext}

where the reduced matrix element $\bra{\lrpy{\la_3}}|\widehat{T}^{\lrpy{\la_2}}|\ket{\lrpy{\la_1}}_{\ga}$ only depends on the representations involved  and the sum is performed over $\ga$, being $\ga$ the index which labels the copies of the $\lrpy{\la_3}$ representation in the $\lrpy{\la_1}\otimes\lrpy{\la_2}$ reduction. We also have that the coupling factor in the expression (\ref{b6}) corresponds to the $q$-deformed Clebsch-Gordan Coefficient of $SU_q(3)$ \cite{Klimyk}.

Moreover, in the $\ket{\psi^{\lrpy{\la}}_{\al}}=\ket{\lrpy{\la},y,t,t_z}$ basis, we have:

\emp
\bra{\lrpy{\la}y't't'_{z}}\widehat{T}^{\lrpy{\nu}}_{\mu}\ket{\lrpy{\la}ytt_z}=\delta^{y'}_{y}\delta^{t'}_{t}\delta^{t'_{z}}_{t_z}\bra{\lrpy{\la}}|\widehat{T}^{\lrpy{\nu}}_{\mu}|\ket{\lrpy{\la}}
\label{b7}
\fin

Therefore, by comparing the previous result with the corresponding to the $q$-deformed Wigner-Eckart theorem, we obtain that $\mu$ should be $\lrp{ytt_z}=\lrp{000}$.\\

Then, according to the previous we have that the $\widehat{M}_T$ operator is given by:

\emp
\widehat{M}_T=\sum_{\nu}\widehat{T}^{\lrpy{\nu}}_{000}
\label{b8}
\fin

where the sum is performed over all physically allowed $SU_q(3)$ representations. Hence, we have $\lrpy{\nu}=\lrpy{0},\lrpy{21},\lrpy{42},\cdots$.\\

Therefore, the particle mass corresponding to a $\ket{\psi^{\lrpy{\la}}_{\al}}=\ket{\lrpy{\la},y,t,t_z}$ state of a multiplet belonging to a $\lrpy{\la}$ representation of $SU_q(3)$ is given by:

\begin{eqnarray}
m\lrp{\lrpy{\la},y,t,t_z}&=&m_S\lrp{\lrpy{\la}}\nonumber\\
&+&\sum_{\nu}\bra{\lrpy{\la},y,t,t_z}\widehat{T}^{\lrpy{\nu}}_{000}\ket{\lrpy{\la},y,t,t_z}\nonumber\\
\label{b9}
\end{eqnarray}

It is important to point out that the $\bra{\psi^{\lrpy{\la}}_{\al}}\widehat{T}^{\lrpy{0}}_{000}\ket{\psi^{\lrpy{\la}}_{\al}}$ term remains absorbed in the $m_S\lrp{\lrpy{\la}}$ term, which is a $SU_q(3)$ scalar. We also have that the dominant contribution comes from the $\widehat{T}^{\lrpy{21}}_{000}$ component of the octet tensor operator. Hence, according to the previous we obtain \cite{Carcamo}:
\begin{eqnarray}
m\lrp{\lrpy{\la},y,t,t_z}&=&m_S\lrp{\lrpy{\la}}\nonumber\\
&+&\bra{\lrpy{\la},y,t,t_z}\widehat{T}^{\lrpy{21}}_{000}\ket{\lrpy{\la},y,t,t_z}\nonumber\\
\label{b10}
\end{eqnarray}

On the other hand, it is well known that the product of the $SU_q\lrp{3}$ representations $\lrpy{21}$ and $\lrpy{21}$ is given by:

\emp
\lrpy{21}\times\lrpy{21}=\lrpy{42}+\lrpy{3}+\lrpy{3^2}+\lrpy{21}_S+\lrpy{21}_A+\lrpy{0}
\label{b11}
\fin

where $\lrpy{42}$, $\lrpy{3}$, $\lrpy{3^2}$, $\lrpy{21}$, and $\lrpy{0}$ are representations with dimensions equal to $27$, $10$, $10$, $8$, and $1$, respectively. Moreover, $\lrpy{21}_S$ and $\lrpy{21}_A$ are symmetric and antisymmetric representations with the same transformation properties under the quantum group $SU_q(3)$.\\
  
We also have that the $\lrpy{21}$ representation corresponding to the octet baryon exhibits the following descomposition under the subgroup $U(1)_Y\times SU_q(2)_T$:

\emp
\lrpy{21}\downarrow\lrpy{3}\times\lrpy{1}+\lrpy{0}\times\lrpy{2}+\lrpy{0}\times\lrpy{0}+\lrpy{\overline{3}}\times\lrpy{1}
\label{b12}
\fin
 
where the products $\lrpy{3}\times\lrpy{1}$, $\lrpy{0}\times\lrpy{2}$, $\lrpy{0}\times\lrpy{0}$, and $\lrpy{\overline{3}}\times\lrpy{1}$ represent $2$, $3$, $1$, and $2$ states corresponding to the $N$, $\Sigma$, $\Lambda$, and $\Xi$ particles of the octet baryon \cite{Haase}.\\

Then, by applying the $q$-deformed Wigner-Eckart theo-rem to the second term of expression (\ref{b10}) taking into account that in the $\lrpy{21}\times\lrpy{21}$ reduction, the $\lrpy{3}\times\lrpy{1}$, $\lrpy{0}\times\lrpy{2}$, $\lrpy{0}\times\lrpy{0}$, and $\lrpy{\overline{3}}\times\lrpy{1}$ states of the first $\lrpy{21}$ representation are coupled to the $\lrpy{0}\times\lrpy{0}$ state of the second one $\lrpy{21}$, where the coupling coefficients are the $q$-deformed isoscalar factors, we obtain the following expressions corresponding to the masses of the particle families in the octet baryon \cite{Carcamo}:

\begin{widetext}  
\begin{eqnarray}
m_N=m_S\lrp{\lrpy{21}}+\begin{pmatrix}
\lrpy{21} & \lrpy{21} & || & \lrpy{21}_S \\
\frac1{2}1 & 00 & || & \frac1{2}1
\end{pmatrix}_q\bra{\lrpy{21}}|\widehat{T}^{\lrpy{21}}|\ket{\lrpy{21}}_S+\begin{pmatrix}\lrpy{21} & \lrpy{21} & || & \lrpy{21}_A \\
\frac1{2}1 & 00 & \left|\right| & \frac1{2}1
\end{pmatrix}_q\bra{\lrpy{21}}|\widehat{T}^{\lrpy{21}}|\ket{\lrpy{21}}_A\nonumber\\
\label{o1}
\end{eqnarray}
\begin{eqnarray}
m_{\Sigma}=m_S\lrp{\lrpy{21}}+\begin{pmatrix}
\lrpy{21} & \lrpy{21} & || & \lrpy{21}_S \\
10 & 00 & || & 10
\end{pmatrix}_q\bra{\lrpy{21}}|\widehat{T}^{\lrpy{21}}|\ket{\lrpy{21}}_S+\begin{pmatrix}
\lrpy{21} & \lrpy{21} & || & \lrpy{21}_A \\
10 & 00 & || & 10
\end{pmatrix}_q\bra{\lrpy{21}}|\widehat{T}^{\lrpy{21}}|\ket{\lrpy{21}}_A\nonumber\\
\label{o2}
\end{eqnarray}
\begin{eqnarray}
m_{\Lambda}=m_S\lrp{\lrpy{21}}+\begin{pmatrix}
\lrpy{21} & \lrpy{21} & || & \lrpy{21}_S \\
00 & 00 & || & 00
\end{pmatrix}_q\bra{\lrpy{21}}|\widehat{T}^{\lrpy{21}}|\ket{\lrpy{21}}_S+\begin{pmatrix}
\lrpy{21} & \lrpy{21} & || & \lrpy{21}_A \\
00 & 00 & || & 00\hspace{0.6cm}
\end{pmatrix}_q\bra{\lrpy{21}}|\widehat{T}^{\lrpy{21}}|\ket{\lrpy{21}}_A\nonumber\\
\label{o3}
\end{eqnarray}
\begin{eqnarray}
m_{\Xi}=m_S\lrp{\lrpy{21}}+\begin{pmatrix}
\lrpy{21} & \lrpy{21} & || & \lrpy{21}_S \\
\frac1{2}-1 & 00 & || & \hspace{0.0cm}\frac1{2}-1
\end{pmatrix}_q\bra{\lrpy{21}}|\widehat{T}^{\lrpy{21}}|\ket{\lrpy{21}}_S+\begin{pmatrix}
\lrpy{21} & \lrpy{21} & || & \lrpy{21}_A \\
\frac1{2}-1 & 00 & || & \hspace{0.0cm}\frac1{2}-1
\end{pmatrix}_q\bra{\lrpy{21}}|\widehat{T}^{\lrpy{21}}|\ket{\lrpy{21}}_A\nonumber\\
\label{o4}
\end{eqnarray}
\end{widetext}  

We also have that according to Racah's factorization lemma, the $q$-deformed isoscalar factors for the $SU_q(3)$ quantum algebra are given by \cite{Yu}:

\begin{eqnarray}
\begin{pmatrix}
\la_1 & \la_2 & || & \la_{3\ga}\\
t_1y_1 & t_2y_2 & || & ty
\end{pmatrix}_q&=&\frac{\begin{pmatrix}
\la_1 & \la_2 &| & \la_{3\ga}\\
\mu_1 & \mu_2 & | & \mu_3
\end{pmatrix}_q}{_qC^{t_1t_2t}_{t_{1z}t_{2z}t_z}}\nonumber\\
&=&\frac{_qC^{\lrpy{\la_1}\times\lrpy{\la_2}}_{(\al_1\al_2)\lrpy{\la_3}(\al_3)}}{_qC^{t_1t_2t}_{t_{1z}t_{2z}t_z}}
\label{o5}
\end{eqnarray}

where $_qC^{t_1t_2t}_{t_{1z}t_{2z}t_z}$ is the $q$-deformed Clebsch-Gordan coefficient for $SU_q(2)$. For this case, we have \cite{Biedenharn,Carcamo}: 

\emp
_qC^{\frac1{2}0\frac1{2}}_{\frac1{2}0\frac1{2}}=\hspace{0.5mm}_qC^{101}_{101}=\hspace{0.5mm}_qC^{000}_{000}=1
\label{o6}
\fin

Then, for the octet baryon we get:

\begin{eqnarray}
\begin{pmatrix}
\lrpy{21} & \lrpy{21} & || & \lrpy{21}_{S,A}\\
ty & 00 & || & ty
\end{pmatrix}_q&=&_qC^{\lrpy{21}\times\lrpy{21}}_{(\al 5)\lrpy{21}_{S,A}(\al)}\nonumber\\
\label{o7}
\end{eqnarray}

Therefore, by combining expressions (\ref{o1})-(\ref{o4}) and using the quantum Clebsch-Gordan coefficients corres-ponding to the $\lrpy{21}\times\lrpy{21}$ reduction obtained in \cite{Ma}, the following $q$-deformed mass relation for octet baryons is obtained \cite{Carcamo}: 
 
\begin{widetext}  
\begin{eqnarray}
\lrpr{3}_q\sqrt{\frac{\lrpr{3}_q+\lrpr{2}_q}{\lrpr{4}_q+1}}m_{\Lambda}+m_{\Sigma}\lrp{C\lrp{q}-E\lrp{q}-D\lrp{q}}=m_N\lrpy{q^{-2}E\lrp{q}+\frac{q^2}{2}\lrp{A\lrp{q}+\frac{\lrp{\lrpr{\frac{3}{2}}_q}^2\sqrt{\lrpr{5}_q}}{\lrpr{3}_q\lrpr{\frac{1}{2}}_q\lrpr{\frac{5}{2}}_q}B\lrp{q}}}\nonumber\\
\hspace{2.5cm}+m_{\Xi}\lrpy{q^{2}E\lrp{q}+\frac{q^3}{2}\lrp{A\lrp{q}-\frac{\lrp{\lrpr{\frac{3}{2}}_q}^2\sqrt{\lrpr{5}_q}}{\lrpr{3}_q\lrpr{\frac{1}{2}}_q\lrpr{\frac{5}{2}}_q}B\lrp{q}}}
\label{p0}
\end{eqnarray}
\end{widetext}

where $A\lrp{q}$, $B\lrp{q}$, $C\lrp{q}$ and $D\lrp{q}$ are functions of the $q$-real parameter given by: 

\begin{widetext}  
\begin{eqnarray}
A\lrp{q}&=&1-q^{-1}+q^{-2}+q^{-3}-q^{-4}+q^{-5},\hspace{0.75cm}B\lrp{q}=1+q^{-1}+q^{-2}-q^{-3}-q^{-4}-q^{-5},\hspace{0.75cm}E\lrp{q}=\sqrt{\frac{\lrpr{3}_q+\lrpr{2}_q}{\lrpr{4}_q+1}}\nonumber\\
C\lrp{q}&=&\frac{q^{5/2}\lrp{\lrpr{2}_q+1}\lrp{q^{3/2}+q^{-3/2}}}{2\lrpr{3}_q}A\lrp{q},\hspace{1cm}D\lrp{q}=\frac{q^{5/2}\lrp{\lrpr{2}_q-1}\lrp{q^{3/2}-q^{-3/2}}\lrp{\lrpr{\frac{3}{2}}_q}^{2}\sqrt{\lrpr{5}_q}}{2\lrp{\lrpr{3}_q}^2\lrpr{\frac{1}{2}}_q\lrpr{\frac{5}{2}}_q}B\lrp{q}
\label{p1}
\end{eqnarray}
\end{widetext}

By replacing the average multiplet masses \cite{PDG} in the expression (\ref{p0}) and performing a program in C that finds the roots of this expression, we obtain \footnotemark[2]:

\emp
q_1=0.9870\pm 0.0002
\label{p2}
\fin

Throughout this article, the errors of the $q$-parameters were obtained by applying the formula:

\emp
\Delta q=\sqrt{\sum\limits^N_{i=1}\lrp{\depar{}{q}{m_i}}^2\lrp{\Delta m_i}^2}
\fin

where $N$ is the number of particles under consideration, $\Delta m_i$ is the error of their masses and $\depar{}{q}{m_{\Lambda}}$, $\depar{}{q}{m_{\Sigma}}$, $\depar{}{q}{m_{N}}$, $\depar{}{q}{m_{\Xi}}$, obtained by performing an implicit differentiation of expression (\ref{p0}) are respectively given by:

\begin{widetext}  
\begin{eqnarray}
\depar{}{q}{m_{\Lambda}}&=&\frac{-F\lrp{q}}{m_{\Lambda}\frac{dF\lrp{q}}{dq}+m_{\Sigma}\frac{dG\lrp{q}}{dq}-m_N\frac{dH\lrp{q}}{dq}-m_{\Xi}\frac{dI\lrp{q}}{dq}},\hspace{1.2cm}\depar{}{q}{m_{\Sigma}}=\frac{-G\lrp{q}}{m_{\Lambda}\frac{dF\lrp{q}}{dq}+m_{\Sigma}\frac{dG\lrp{q}}{dq}-m_N\frac{dH\lrp{q}}{dq}-m_{\Xi}\frac{dI\lrp{q}}{dq}}\nonumber\\
\depar{}{q}{m_{N}}&=&\frac{H\lrp{q}}{m_{\Lambda}\frac{dF\lrp{q}}{dq}+m_{\Sigma}\frac{dG\lrp{q}}{dq}-m_N\frac{dH\lrp{q}}{dq}-m_{\Xi}\frac{dI\lrp{q}}{dq}},\hspace{1.2cm}\depar{}{q}{m_{\Xi}}=\frac{I\lrp{q}}{m_{\Lambda}\frac{dF\lrp{q}}{dq}+m_{\Sigma}\frac{dG\lrp{q}}{dq}-m_N\frac{dH\lrp{q}}{dq}-m_{\Xi}\frac{dI\lrp{q}}{dq}}\nonumber
\end{eqnarray}
\end{widetext}

\footnotetext[2]{In this article we choose between the roots of the mass relations which satisfy $0\leqslant q\leqslant 1$.}
\vspace{-3cm}
With the functions $F\lrp{q}$, $G\lrp{q}$, $H\lrp{q}$ and $I\lrp{q}$ given by:

\begin{eqnarray}
F\lrp{q}&=&\F\nonumber
\end{eqnarray}
\begin{eqnarray}
G\lrp{q}&=&\G\nonumber
\end{eqnarray}
\begin{eqnarray}
H\lrp{q}&=&\Ha\nonumber
\end{eqnarray}
\begin{eqnarray}
I\lrp{q}&=&\I\nonumber
\end{eqnarray}

To determine the $q$-deformed mass relations for the decuplet baryon, we follow the same procedure used for the octet baryon, taking into account that when the $q$-deformed Wigner Eckart theorem is applied to the se-cond term of expression (\ref{b10}), the $\lrpy{3}\times\lrpy{3}$, $\lrpy{0}\times\lrpy{2}$, $\lrpy{\overline{3}}\times\lrpy{1}$, and $\lrpy{\overline{6}}\times\lrpy{0}$ states of the $\lrpy{3}$ representation (which represent $4$, $3$, $2$, and $1$ states respectively, corresponding to the $\Delta$, $\Sigma$, $\Xi$, and $\Omega$ particles of the decuplet baryon) are coupled to the $\lrpy{0}\times\lrpy{0}$ state of the $\lrpy{21}$ representation, where the coupling coefficients are the $q$-deformed isoscalar factors. Hence, according to the previous, we obtain the following expressions corresponding to the masses of the particle families in the decuplet baryon \cite{Carcamo}:

\begin{widetext}  
\begin{eqnarray}
m_{\Delta}=m_S\lrp{\lrpy{3}}+\begin{pmatrix}
\lrpy{3} & \lrpy{21} & || & \lrpy{3}\\
\frac{3}{2}1 & 00 & || & \frac{3}{2}1
\end{pmatrix}_q\bra{\lrpy{3}}|\widehat{T}^{\lrpy{21}}|\ket{\lrpy{3}}
\label{g04}
\end{eqnarray}
\begin{eqnarray}
m_{\Sigma}=m_S\lrp{\lrpy{3}}+\begin{pmatrix}
\lrpy{3} & \lrpy{21} & || & \lrpy{3}\\
10 & 00 & || & 10
\end{pmatrix}_q\bra{\lrpy{3}}|\widehat{T}^{\lrpy{21}}|\ket{\lrpy{3}}
\label{g5}
\end{eqnarray}
\begin{eqnarray}
m_{\Xi}=m_S\lrp{\lrpy{3}}+\begin{pmatrix}
\lrpy{3} & \lrpy{21} & || & \lrpy{3}\\
\frac1{2}-1 & 00 & || & \frac1{2}-1
\end{pmatrix}_q\bra{\lrpy{3}}|\widehat{T}^{\lrpy{21}}|\ket{\lrpy{3}}
\label{g6}
\end{eqnarray}
\begin{eqnarray}
m_{\Omega}=m_S\lrp{\lrpy{3}}+\begin{pmatrix}
\lrpy{3} & \lrpy{21} & || & \lrpy{3}\\
0-2 & 00 & || & \hspace{0.0cm}0-2
\end{pmatrix}_q\bra{\lrpy{3}}|\widehat{T}^{\lrpy{21}}|\ket{\lrpy{3}}
\label{g7}
\end{eqnarray}
\end{widetext}  

Where for this case, we have \cite{Biedenharn,Carcamo}:

\emp
_qC^{\frac{3}{2}0\frac{3}{2}}_{\frac{3}{2}0\frac{3}{2}}=\hspace{0.5mm}_qC^{101}_{101}=\hspace{0.5mm}_qC^{\frac{1}{2}0\frac{1}{2}}_{\frac{1}{2}0\frac{1}{2}}=1
\label{g8}
\fin

Then, for the decuplet baryon we have:

\begin{eqnarray}
\begin{pmatrix}
\lrpy{3} & \lrpy{21} & || & \lrpy{3}\\
ty & 00 & || & ty
\end{pmatrix}_q&=&_qC^{\lrpy{3}\times\lrpy{21}}_{(\al 5)\lrpy{3}(\al)}\nonumber\\
\label{g9}
\end{eqnarray}

Therefore, by combining expressions (\ref{g04})-(\ref{g7}) and using the $q$-deformed Clebsch-Gordan coefficients corres-ponding to the $\lrpy{3}\times\lrpy{21}$ reduction given in the Appendix, the following $q$-deformed mass relations for decuplet baryons is obtained \cite{Carcamo}: 

\begin{eqnarray}
m_{\Sigma^*}-m_{\Omega}=\frac{1+q}{\lrpr{2}_q}\lrp{m_{\Delta}-m_{\Xi^*}}\label{g300}
\end{eqnarray}
\begin{eqnarray}
m_{\Xi^*}-m_{\Omega}=\frac{q^3}{1+q}\lrp{m_{\Sigma^*}-m_{\Omega}}\label{g101}
\end{eqnarray}
\begin{eqnarray}
m_{\Omega}-m_{\Xi^*}=\frac{q^3}{1+q-q^3}\lrp{m_{\Xi^*}-m_{\Sigma^*}}\label{g102}\end{eqnarray}

where the $q$-deformation parameter, according to the expression (\ref{g300}) is given by: 

\begin{widetext}  
\begin{eqnarray}
q=\frac{m_{\Delta}-m_{\Xi^*}\pm\sqrt{\lrp{m_{\Delta}-m_{\Xi^*}}^2-4\lrp{m_{\Sigma^*}-m_{\Omega}}\lrp{m_{\Sigma^*}+m_{\Xi^*}-m_{\Omega}-m_{\Delta}}}}{2\lrp{m_{\Sigma^*}+m_{\Xi^*}-m_{\Omega}-m_{\Delta}}}
\label{g4}
\end{eqnarray}
\end{widetext}

As in the case of the octet baryons, we replace the ave-rage multiplet masses obtaining the following values for the $q$-parameter:

\begin{eqnarray}
q_2&=&0.917\pm 0.012\hspace{1.2cm}q_3=0.986\pm 0.003\nonumber\\
q_4&=&0.985\pm 0.002
\label{g6}
\end{eqnarray}

With the aim to determine a unique $q$-parameter for the $q$-deformed Gell-Mann-Okubo mass relations for octet and decuplet baryons, we perform a $\chi^2$ adjustment given by the fitted function:

\emp
\chi^2=\sum\limits^{4}_{i=1}\frac{\lrp{q_i-q}^2}{q^2}
\label{e26}
\fin

By minimizing the $\chi^2$ function, we obtain that the fitted $q$-parameter is given by:
\emp
q=0,970\pm 0.003
\label{e280}
\fin
    
Besides that, by comparing the generalization of the Gell-Mann-Okubo mass formula for pseudoescalar meson with its $q$-deformed version, the following relation is obtained \cite{Gavrilik}:

\emp
\frac{f^2_K}{f^2_{\pi}}=\frac{\lrpr{2}_q}{2\lrp{2\lrpr{2}_q-\lrpr{3}_q}}
\label{30e}
\fin

where $f_K$ and $f_{\pi}$ are the decay constants for the $K$ and $\pi$ mesons, respectively.\\

Moreover, the ratio $\frac{f_K}{f_{\pi}}$ can be expressed in terms of the Cabbibo angle as follows \cite{Gavrilik}:

\emp
\tan^2\theta_C=\frac{\frac{m^2_{\pi}}{m^2_K}}{\frac{f_K}{f_{\pi}}-\frac{m^2_{\pi}}{m^2_K}}
\label{30f}
\fin 

From expressions (\ref{30e}) and (\ref{30f}), a connection between the $q$-deformation parameter and the Cabbibo angle is observed \cite{Gavrilik}.\\

On the other hand, as we have chosen before that $0\leqslant q\leqslant 1$, we can introduce a new $\tau$ parameter according to:
\emp
q=\cos\tau
\label{e29}
\fin

Then, from expressions (\ref{e280}) and (\ref{e29}) we get:
\emp
\tau=\lrp{0.246\pm 0.012}rad
\label{e30c}
\fin

As the Cabbibo angle is equal to $\theta_C=\lrp{0.226\pm 0.002}$rad \cite{PDG}, the expression (\ref{e30c}) implies:

\emp
\tau=\lrp{1.085\pm 0.063}\theta_C
\label{e31}
\fin
Hence, according to the previous result, it is possible to interpretate the $q$-deformation parameter with the cosine of the Cabbibo angle.\\

By replacing the $q$-deformation parameter obtained by the $\chi^2$ fitting in the Gell-Mann-Okubo mass relations and comparing with the $q=1$ case, we obtain Table \ref{tab:table1}.  
\begin{table}
\caption{\label{tab:table1}Error percentages of the Gell-Mann-Okubo mass relations for the $SU_q(3)$ and $SU(3)$ algebras.}
\begin{tabular}{ccc}
\hline
Mass Relation & $q$ & $\frac{|RHS-LHS|}{RHS}\%$\\
\hline
(\ref{p0}) & 0.970 & 0.33\\ 
(\ref{p0}) & 1.000 & 0.58\\
(\ref{g300}) & 0.970 & 2.99\\
(\ref{g300}) & 1.000 & 4.49\\
(\ref{g101}) & 0.970 & 4.26\\
(\ref{g101}) & 1.000 & 3.40\\
(\ref{g102}) & 0.970 & 7.95\\
(\ref{g102}) & 1.000 & 8.23\\
\hline
\end{tabular}
\end{table}
In Table I we can see that for the $q$-deformed Gell-Mann-Okubo mass relations for the octet and decuplet baryons, a better agreement with the experimental masses than the predicted by the $SU(3)$ algebra is obtained, except for expression (\ref{g101}) corresponding to the decuplet baryon, where the difference between the error percentages is only $0.86\%$.

\section{The $q$-deformed Coleman-Glashow relations for octet and decuplet baryons}
The center of the octet baryon is degenerate and so the $U$ eigenfunctions differ from isoespin eigenfunctions. The $U=1$ and $U=0$ eigenfuntions are:

\begin{eqnarray}
\ket{\Sigma^0_U}=\frac{\sqrt{\lrpr{3}_q}}{\lrpr{2}_q}\ket{\Lambda^0}-\frac1{\lrpr{2}_q}\ket{\Sigma^0}
\label{e0}
\end{eqnarray}
\begin{eqnarray}
\ket{\Lambda^0_U}=\frac{\sqrt{\lrpr{3}_q}}{\lrpr{2}_q}\ket{\Lambda^0}+\frac1{\lrpr{2}_q}\ket{\Sigma^0}
\label{e1}
\end{eqnarray}

For this case, we consider the following expression for the mass operator:

\emp
\widehat{M}=\widehat{M}_S+\widehat{M}_T+\widehat{M}_U
\label{e0}
\fin

where the operators $\widehat{M}_S$, $\widehat{M}_T$ and $\widehat{M}_U$ are $SU_q(3)$ scalar, isospin scalar and $U$ spin scalar, respectively.

The mass operators $\widehat{M}_T$ and $\widehat{M}_U$ have matrix elements related by:

\begin{widetext}
\begin{eqnarray}
m_T\lrp{n}&=&m_T\lrp{p},\hspace{1cm}m_T\lrp{\Sigma^+}=m_T\lrp{\Sigma^0},\hspace{1cm}m_U\lrp{n}=m_U\lrp{\Sigma^0_U},\hspace{1cm}m_U\lrp{p}=m_U\lrp{\Sigma^+_U}\nonumber\\
m_T\lrp{\Sigma^-}&=&m_T\lrp{\Sigma^0},\hspace{0.8cm}m_T\lrp{\Xi^-}=m_T\lrp{\Xi^0},\hspace{0.8cm}m_U\lrp{\Sigma^-}=m_U\lrp{\Xi^-},\hspace{0.8cm}m_U\lrp{\Xi^0}=m_U\lrp{\Sigma^0_U}
\label{e4}
\end{eqnarray}
\end{widetext}

Moreover, by taking into account that:

\empn
m_U\lrp{\Sigma^0_U}=\bra{\Sigma^0_U}\widehat{M}_U\ket{\Sigma^0_U}\hspace{0.6cm}\bra{\Sigma^0_U}\widehat{M}_U\ket{\Lambda^0_U}=0
\label{e5}
\finn 
\empn
m\lrp{\Sigma^0\Lambda^0}=\bra{\Sigma^0}\widehat{M}\ket{\Lambda^0}=\bra{\Sigma^0}\widehat{M}_U\ket{\Lambda^0}=m_U\lrp{\Sigma^0\Lambda^0}
\label{e6}
\finn

and using the expressions corresponding to the $\ket{\Sigma^0_U}$, $\ket{\Lambda^0_U}$ states, we get:

\emp
m_U\lrp{\Sigma^0_U}=\frac{\lrpr{3}_qm_U\lrp{\Lambda^0}+m_U\lrp{\Sigma^0}-2\sqrt{\lrpr{3}_q}m_U\lrp{\Sigma^0\Lambda^0}}{\lrp{\lrpr{2}_q}^2}
\label{e7}
\fin
\emp
m_U\lrp{\Lambda^0}-m_U\lrp{\Sigma^0}=-\frac{\lrpr{4}_q}{\lrpr{2}_q\sqrt{\lrpr{3}_q}}m\lrp{\Sigma^0\Lambda^0}
\label{e8}
\fin

Then, from the expressions (\ref{e0})-(\ref{e7}), we find:

\begin{widetext}
\begin{eqnarray}
m\lrp{n}-m\lrp{p}+m\lrp{\Sigma^+}-m\lrp{\Sigma^0}=\frac{\lrpr{3}_q}{\lrp{\lrpr{2}_q}^2}\lrp{m_U\lrp{\Lambda^0}-m_U\lrp{\Sigma^0}}-\frac{2\sqrt{\lrpr{3}_q}}{\lrp{\lrpr{2}_q}^2}m_U\lrp{\Sigma^0\Lambda^0}
\label{6}
\end{eqnarray}
\begin{eqnarray}
m\lrp{\Xi^-}-m\lrp{\Xi^0}+m\lrp{\Sigma^0}-m\lrp{\Sigma^-}=\frac{2\sqrt{\lrpr{3}_q}}{\lrp{\lrpr{2}_q}^2}m_U\lrp{\Sigma^0\Lambda^0}+\frac{\lrpr{3}_q}{\lrp{\lrpr{2}_q}^2}\lrp{m_U\lrp{\Sigma^0}-m_U\lrp{\Lambda^0}}
\label{01}
\end{eqnarray}
\end{widetext}

Therefore, by replacing (\ref{e8}) in expressions (\ref{6}) and (\ref{01}), the following relations are obtained:

\begin{eqnarray}
m_n-m_p+m_{\Sigma^+}-m_{\Sigma^0}=-\sqrt{\lrpr{3}_q}m_{\Sigma^0\Lambda^0}\\
m_{\Xi^-}-m_{\Xi^0}+m_{\Sigma^0}-m_{\Sigma^-}=\sqrt{\lrpr{3}_q}m_{\Sigma^0\Lambda^0}
\label{e9}
\end{eqnarray}

From the linear combinations of the previous relations we find:
\emp
m_{\Xi^-}-m_{\Xi^0}=m_{\Sigma^-}-m_{\Sigma^+}+m_p-m_n
\label{e10}
\fin
Hence, the Colemann-Glashow relation for the octet baryon is independent of $q$.\\

On the other hand, it is known that when the electromagnetic interactions are neglected, we have that the mass operator for hadron multiplets is given by:
\emp
\widehat{M}=\widehat{M}_S+\widehat{M}_T
\label{e101}
\fin
where $\widehat{M}_S$ is a $SU_q(3)$ invariant whereas $\widehat{M}_T$ is an ope-rator corresponding to the $SU_q(3)$ symmetry breaking.\\

We also have that under $SU_q(2)_U\times U(1)_Q$ the $\widehat{M}_T$ operator in expression (\ref{e101}) transforms as the sum of a $U$ vector spin and a $U$ scalar \cite{Carruthers}. Applying the $SU_q(2)$ Wigner-Eckart theorem to this subgroup, we have:

\emp
M\lrp{U,U_3}=A+C\lrp{U,q}q^{-U_3}U_3
\label{e11}
\fin
where $C\lrp{U,q}$ is given by \cite{Yu}:

\emp
C\lrp{U,q}=\frac{q^{U+\frac1{2}}}{\lrpr{2U+1}_q}\bra{\al U}|\widehat{M}_U|\ket{\al U}
\label{e12}
\fin

From expressions (\ref{e11}) and (\ref{e12}) we get:

\begin{eqnarray}
m\lrp{U,U_3}-m\lrp{U,U_3-1}=C\lrp{U,q}q^{-U_3+1}\hspace{1cm}\nonumber\\
+C\lrp{U,q}q^{-U_3}\lrp{1-q}U_3
\label{e13}
\end{eqnarray}

By applying the previous formula to the octet baryon, we find:

\begin{eqnarray*}
m\lrp{n}-m\lrp{\Sigma^0_U}=C\lrp{1,q}q^{-1}
\end{eqnarray*}
\begin{eqnarray*}
m\lrp{\Sigma^0_U}-m\lrp{\Xi^0}=C\lrp{1,q}q
\label{e14}
\end{eqnarray*}
Then, the following relation holds:

\emp
q\lrp{m_n-m_{\Sigma^0_U}}=q^{-1}\lrp{m_{\Sigma^0_U}-m_{\Xi^0}}
\label{e15}
\fin

Hence, by using (\ref{e7}), (\ref{e8}), and (\ref{e15}) we obtain a $q$ deformed mass relation for the particles in the octet baryon:

\emp
q^2m_{n}+m_p+q^{-2}m_{\Xi^0}+m_{\Xi^-}=\lrpr{3}_qm_{\Lambda^0}+m_{\Sigma^+}+m_{\Sigma^-}-m_{\Sigma^0}
\label{e16}
\fin

with the $q$ deformation parameter given by:

\emp
q=\pm\sqrt{\frac{s \pm\sqrt{s^2-4\lrp{m_n-m_{\Lambda^0}}\lrp{m_{\Xi^0}-m_{\Lambda^0}}}}{2\lrp{m_n-m_{\Lambda^0}}}}
\label{e17}
\fin

where the $s$ parameter has been introduced according to: 

\emp
s=m_{\Sigma^+}+m_{\Sigma^-}+m_{\Lambda^0}-m_p-m_{\Xi^-}-m_{\Sigma^0}
\label{e18}
\fin

By replacing the experimental octet particle masses in expression (\ref{e17}), we obtain: 

\emp
q_1=0.965\pm 0.001
\label{e19}
\fin

On the other hand, when formula (\ref{e13}) is applied to the decuplet baryon, we get the following expressions:

\begin{eqnarray*}
m_{\Delta^-}-m_{\Sigma^{*-}}=\frac1{2}C\lrp{\frac{3}{2},q}q^{-3/2}\lrp{3-q}\label{e200}
\end{eqnarray*}
\begin{eqnarray*}
m_{\Sigma^{*-}}-m_{\Xi^{*-}}=\frac1{2}C\lrp{\frac{3}{2},q}q^{-1/2}\lrp{1+q}\label{e201}
\end{eqnarray*}
\begin{eqnarray*}
m_{\Xi^{*-}}-m_{\Omega^{-}}=\frac1{2}C\lrp{\frac{3}{2},q}q^{1/2}\lrp{3q-1}\label{e202}
\end{eqnarray*}
\begin{eqnarray*}
m_{\Delta^0}-m_{\Sigma^{*0}}=C\lrp{1,q}q^{-1}
\end{eqnarray*}
\begin{eqnarray*}
m_{\Sigma^{*0}}-m_{\Xi^{*0}}=C\lrp{1,q}q
\label{e203}
\end{eqnarray*}

Hence, for the decuplet baryons we obtain:

\begin{eqnarray}
\frac{q^{3/2}\lrp{m_{\Delta^-}-m_{\Sigma^{*-}}}}{3-q^{-1}}=\frac{q^{1/2}\lrp{m_{\Sigma^{*-}}-m_{\Xi^{*-}}}}{1+q}\label{e230}\nonumber\\
=\frac{q^{-1/2}\lrp{m_{\Xi^{*-}}-m_{\Omega^{-}}}}{3q-1}
\label{e23}
\end{eqnarray}
\begin{eqnarray}
q\lrp{m_{\Delta^0}-m_{\Sigma^{*0}}}=q^{-1}\lrp{m_{\Sigma^{*0}}-m_{\Xi^{*0}}}
\label{e24}
\end{eqnarray}

Then, from expressions (\ref{e23}) and (\ref{e24}) and using the experimental decuplet particle masses we obtain:

\begin{eqnarray}
q_2=0.976\pm 0,007\hspace{1.2cm}q_3=0,965\pm 0,004\\
q_4=0.970\pm 0,003\hspace{1.2cm}q_{5}=0,988\pm 0,008
\label{e25}
\end{eqnarray}

Up to this point we have obtained five different values for the $q$-deformation parameter from the $q$-deformed Coleman-Glashow relations for octet and decuplet baryons. In order to obtain a unique $q$-parameter we perform a $\chi^2$ adjustment obtaining:
\emp
q=0,973\pm 0.002
\label{e28}
\fin

For this case, we have obtained that the $\tau$ parameter is given by:
\emp
\tau=\lrp{0.233\pm 0.010}rad=\lrp{1.030\pm 0.053}\theta_C
\label{e30}
\fin

which implies that the $q$-deformation parameter can be interpreted with the cosine of the Cabbibo angle.\\ 
By replacing the $q$-deformation parameter obtained by the $\chi^2$ adjustment in the Coleman-Glashow relations and comparing with the predicted by the $SU(3)$ algebra, we obtain the Table \ref{tab:table2}.
\begin{table}
\caption{\label{tab:table2}Error percentages of the Coleman-Glashow relations for the $SU_q(3)$ and $SU(3)$ algebras.}
\begin{tabular}{ccc}
\hline
Mass Relation & $q$ & $\frac{|RHS-LHS|}{RHS}\%$\\
\hline
(\ref{e16}) & 0.973 & 0.14\\ 
(\ref{e16}) & 1.000 & 0.60\\
(\ref{e23}) & 0.973 & 1.29\\
(\ref{e23}) & 1.000 & 1.42\\
(\ref{e23}) & 0.973 & 1.27\\
(\ref{e23}) & 1.000 & 7.03\\
(\ref{e23}) & 0.973 & 1.03\\
(\ref{e23}) & 1.000 & 9.72\\
(\ref{e24}) & 0.973 & 2.97\\
(\ref{e24}) & 1.000 & 2.51\\
\hline
\end{tabular}
\end{table}
From Table II, a better agreement of the $q$-deformed Coleman-Glashow relations with the experiment than the predicted by the $SU(3)$ algebra is obtained in all cases except in the last (relation (\ref{e24})), where the difference between the error percentages is only $0.46\%$.

\section{Conclusions}
The quantum group $SU_q(3)$ provides a good tool to solve problems in particle physics, especially when one needs to describe the mass splitting for particles from isomultiplets within octet and decuplet baryons.\\ 

The $q$-deformed Clebsch-Gordan coefficients corres-ponding to the $\lrpy{3}\times\lrpy{21}$ reduction of the $SU_q\lrp{3}$ algebra were computed. The $q$-deformed mass relations for octet and decuplet baryons have been explicity obtained from the quantum Clebsch-Gordan coefficients corresponding to the $\lrpy{21}\times\lrpy{21}$ and $\lrpy{3}\times\lrpy{21}$ reductions. The Coleman-Glashow relations for octet and decuplet baryons have been found in the $SU_q(3)$ quantum algebra. From the Coleman-Glashow relation for the octet baryon, a $q$-deformed mass relation between its particles has been obtained.\\ 

By performing an adjustment of the $q$-deformation parameter in the $q$-deformed Gell-Mann-Okubo and Coleman-Glashow relations for octet and decuplet baryons, we obtain that the corresponding values for this parameter are $q=0,970\pm 0.003$ and $q=0,973\pm 0.002$, respectively. These values are directly connected with the cosine of the Cabbibo angle. That is, a unique relation between the $q$-deformation parameter and the Cabbibo angle has been found, which differs with the results given in \cite{Gavrilik} in the fact that in this reference two different relations $q=e^{2i\theta_C}$ and $q=e^{i\theta_C}$, between these parameters have been obtained for the octet and decuplet baryons, respectively, exhibiting arror percentages of approximately $0.07\%$ and $0.53\%$ when these relations are replaced in the $q$ deformed masses expressions.\\

In spite of the fact that the error percentages of the $q$-deformed mass relations obtained in \cite{Gavrilik} are lower than those obtained in this article, it is important to point out that we have shown that when two approximately\hspace{1mm} equi-\\valent values of the $q$-deformation parameter are used, the $q$-deformed Gell-Mann-Okubo and Coleman-Glashow relations for octet and decuplet baryons exhibit a very good agreement with the experimental results, in most cases better than the predicted by the $SU(3)$ algebra. The error percentages of the $q$-deformed Gell-Mann-Okubo and Coleman-Glashow relations are lower than $7.95\%$ and $2.97\%$, respectively.

\section*{ACKNOWLEDGMENTS}
The author thanks Professor Richard W Haase for introducing him to this field, for reviewing the calculations, and for carefully reading the manuscript. The author also thanks Professor R. Hurtado and Professor M. Di. Sanctis for reading the manuscript and giving their suggestions and comments. Moreover, the author is grateful to Mateo Restrepo for facilitating useful references. 

\appendix
\section{Quantum Clebsch-Gordan Coefficients for the $\lrpy{3}\times\lrpy{21}$ reduction }
The product of the $SU_q(3)$ representations $\lrpy{3}$ and $\lrpy{21}$ is given by:

\emp
\lrpy{3}\times\lrpy{21}=\lrpy{51}+\lrpy{42}+\lrpy{3}+\lrpy{21}
\label{ap1}
\fin
where $\lrpy{51}$, $\lrpy{42}$, $\lrpy{3}$, and $\lrpy{21}$ are representations with dimensions equal to $35$, $27$, $10$, and $8$ respectively. These representations are shown in the Figures 1-4.\\

Through a straightforward calculation in which the angular momentum addition rules are taken into account together with the requirements of orthogonality, the $q$-deformed Clebsch-Gordan coefficients corresponding\hspace{1.56mm}to\begin{figure}
\includegraphics[width=7.5cm,height=7.5cm,angle=0]{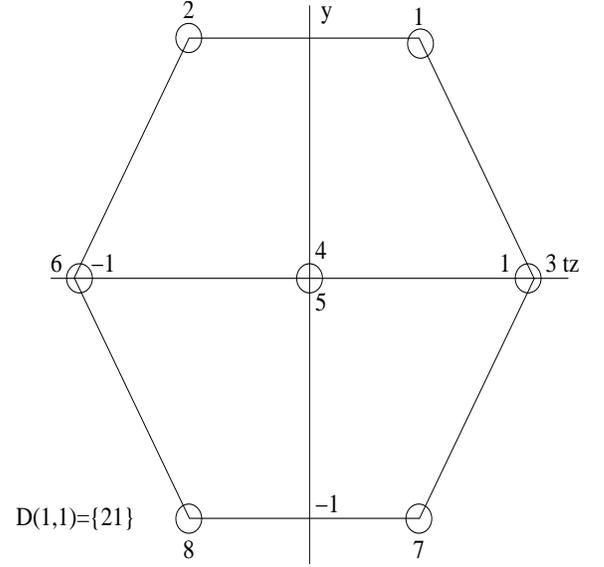}
\caption{\label{fig 1:epsart} Representation $\lrpy{21}$ of $SU_q(3)$ \cite{Carcamo}.}
\end{figure}
\begin{figure}
\includegraphics[width=7cm,height=7.5cm,angle=0]{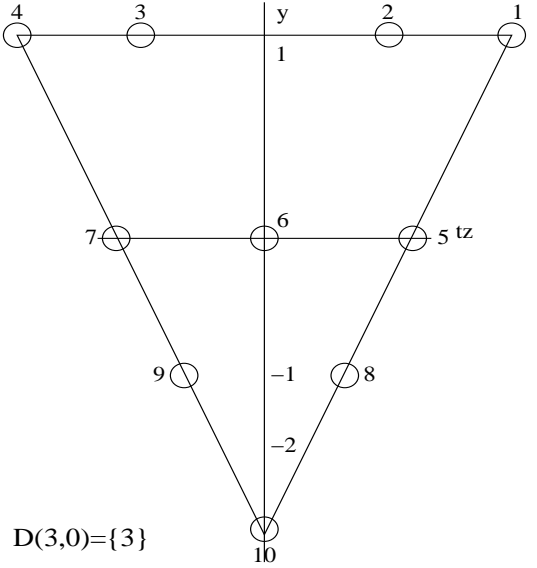}
\caption{\label{fig 2:epsart} Representation $\lrpy{3}$ of $SU_q(3)$ \cite{Carcamo}.}
\end{figure}
\newpage
\begin{widetext}
\begin{center}
\begin{figure}
\includegraphics[width=10cm,height=8.35cm,angle=0]{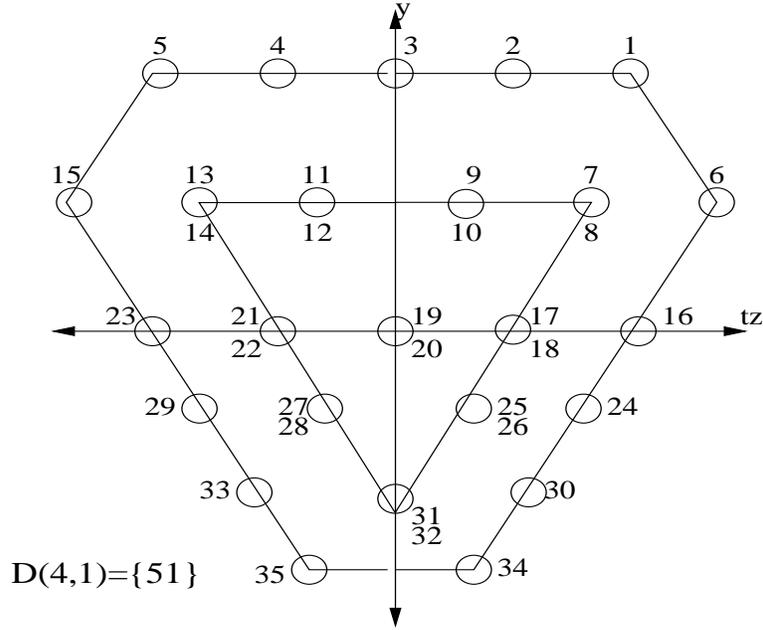}
\caption{\label{fig 3:epsart} Representation $\lrpy{51}$ of $SU_q(3)$ \cite{Carcamo}.}
\end{figure}
\begin{figure}
\includegraphics[width=10cm,height=8.35cm,angle=0]{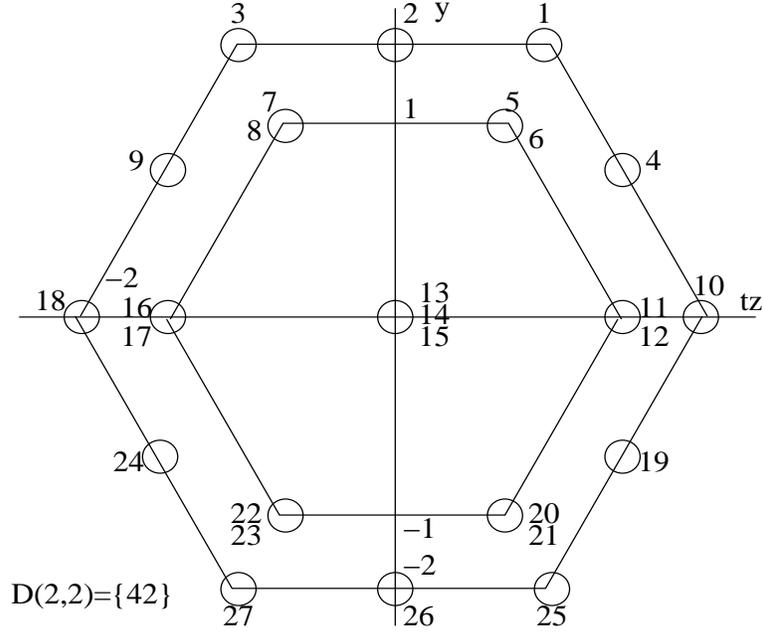}
\caption{\label{fig 4:epsart} Representation $\lrpy{42}$ of $SU_q(3)$ \cite{Carcamo}.}
\end{figure}
\end{center}
\end{widetext}
the $\lrpy{3}\times\lrpy{21}$ reduction are determined and shown in the Table \ref{tab:table3}. In this case, we have that $\chi_i$, $i=1,2\cdots 8$ and $\xi_j$, $j=1,2\cdots 10$ are the wave functions which describe the octet and decuplet baryons states corresponding to the $\lrpy{21}$ and $\lrpy{3}$ representations.
   
\begin{table*}
\caption{\label{tab:table3}The $q$-deformed Clebsch-Gordan coefficients for the $\lrpy{3}\times\lrpy{21}$ reduction.}
\begin{tabular}{cccccccc}
\hline
 & & (a)\0\0\0$\psi^{\lrpy{51}}_1$ & (b)\0\0\0$\psi^{\lrpy{51}}_5$ & (c)\0\0\0$\psi^{\lrpy{51}}_{6}$ & (d)\0\0\0$\psi^{\lrpy{51}}_{15}$ & (e)\0\0\0$\psi^{\lrpy{51}}_{34}$ & (f)\0\0\0$\psi^{\lrpy{51}}_{35}$ \\
\hline
(a)\0 $\xi_1\h_1$ \0\0 & (d)\0 $\xi_4\h_6$ & & & & & & \\
(b)\0 $\xi_4\h_2$ \0\0 & (e)\0 $\xi_{10}\h_7$ & & & \hspace{1cm}1 & & & \\
(c)\0 $\xi_{1}\h_3$ \0\0 & (f)\0 $\xi_{10}\h_8$ & & & & & & \\
\hline
& & (a)\0 $\psi^{\lrpy{51}}_2$\0 & \m$\psi^{\lrpy{42}}_{1}$\hspace{3mm}& (b)\0$\psi^{\lrpy{51}}_4$\0 & \0\0$\psi^{\lrpy{42}}_3$\hspace{6.5mm} & (c)\0 $\psi^{\lrpy{51}}_{3}$\0\0 & \0\hspace{0.5mm}\m $\psi^{\lrpy{42}}_{2}$\\
& & (d)\0 $\psi^{\lrpy{51}}_{16}$\0\0 & \0$\psi^{\lrpy{51}}_{30}$\hspace{0.4cm} & (e)\0$\psi^{\lrpy{42}}_{10}$\0\0 & \0$\psi^{\lrpy{42}}_{25}$\hspace{0.7cm} & (f)\0 $\psi^{\lrpy{51}}_{24}$\0\0 & \0\hspace{0.5mm}$\psi^{\lrpy{42}}_{19}$\\
& & (g)\0 $\psi^{\lrpy{51}}_{23}$\0\0 & \0$\psi^{\lrpy{51}}_{33}$\hspace{0.4cm} & (h)\0$\psi^{\lrpy{42}}_{18}$\0\0 & \0$\psi^{\lrpy{42}}_{27}$\hspace{0.7cm} & (i)\0 $\psi^{\lrpy{51}}_{29}$\0\0 & \0\hspace{0.5mm}$\psi^{\lrpy{42}}_{24}$ \\
\hline
(a)\0 $\xi_2\h_1$ & (b)\0\hspace{0mm}$\xi_3\h_2$ & & & & & & \\
(d)\0 $\xi_5\h_3$ & \0\hspace{1.5mm}(e)\0 $\xi_{8}\h_7$\0\0 & $\sqrt{\frac{q\lrpr{3}_q}{\lrpr{4}_q}}$\0 & $\sqrt{\frac{1}{q^3\lrpr{4}_q}}$\0\0 & $\sqrt{\frac{\lrpr{3}_q}{q\lrpr{4}_q}}$\0\0 & \m$\sqrt{\frac{q^3}{\lrpr{4}_q}}$ & \hspace{0cm}0 & 0 \\
(g)\0 $\xi_{7}\h_6$ & (h)\0\hspace{0mm}$\xi_{9}\h_8$ & & & & & &\\
(c)\0 $\xi_2\h_2$ & (c)\0\hspace{0mm}$\xi_3\h_1$ & & & & & \hspace{0cm}$q^{-1}\sqrt{\frac{\lrpr{2}_q}{\lrpr{4}_q}}$ & \hspace{0.1cm}\m $q\sqrt{\frac{\lrpr{2}_q}{\lrpr{4}_q}}$ \\
(f)\0 $\xi_5\h_7$ & (f)\0 $\xi_{8}\h_3$ & \hspace{0mm}0 & \hspace{0mm}0\hspace{0mm} & 0 & \hspace{0mm}0 & & \\
(c)\0 $\xi_7\h_8$ & (c)\0\hspace{0mm}$\xi_9\h_6$ & & & & & \hspace{0cm}$q\sqrt{\frac{\lrpr{2}_q}{\lrpr{4}_q}}$ & \hspace{0.1cm}$q^{-1}\sqrt{\frac{\lrpr{2}_q}{\lrpr{4}_q}}$ \\
(a)\0 $\xi_1\h_2$ & (b)\hspace{2.5mm}$\xi_4\h_1$ & & & & & & \\
(d)\0 $\xi_1\h_7$ & (e)\hspace{2.5mm}$\xi_{10}\h_3$ & $\sqrt{\frac1{q^{3}\lrpr{4}_q}}$ & \m$\sqrt{\frac{q\lrpr{3}_q}{\lrpr{4}_q}}$ & $\sqrt{\frac{q^3}{\lrpr{4}_q}}$ & $\sqrt{\frac{\lrpr{3}_q}{q\lrpr{4}_q}}$ & \hspace{0.5mm}0 & \hspace{0.5mm}0 \\
(g)\0 $\xi_{4}\h_8$ & (h)\hspace{2.5mm}$\xi_{10}\h_6$ & & & & & & \\
\hline
& & (a)\hspace{0.45cm}$\psi^{\lrpy{51}}_{31}$ & $\psi^{\lrpy{51}}_{32}$ & $\psi^{\lrpy{42}}_{26}$ & $\psi^{\lrpy{3}}_{10}$ & (b)\hspace{0.2cm}$\psi^{\lrpy{51}}_{19}$ & $\psi^{\lrpy{51}}_{20}$\\
\hline 
(a)\0$\xi_{10}\h_4$ & (b)\0$\xi_5\h_6$ & $\sqrt{\frac{q^3}{\lrpr{4}_q}}$ & 0 & $\sqrt{\frac{\lrpr{3}_q}{q\lrpr{4}_q}}$ & 0 & $\frac{\sqrt{q^{-3}\lrpr{2}_q}}{\lrpr{4}_q}$ & $\frac{\lrpr{2}_q}{\lrpr{4}_q}\sqrt{\frac{1}{q^3\lrpr{6}_q}}$\\
& & & & & & & \\
(a)\0$\xi_{10}\h_5$ & (b)\0$\xi_7\h_3$ & 0 & $\sqrt{\frac{q^3\lrpr{3}_q}{\lrpr{6}_q}}$ & 0 & $\sqrt{\frac{\lrpr{3}_q}{q^3\lrpr{6}_q}}$ & $\frac{\sqrt{q^5\lrpr{2}_q}}{\lrpr{4}_q}$ & \m$\frac{\lrpr{2}_q}{\lrpr{4}_q}\sqrt{\frac1{q^3\lrpr{6}_q}}$\\
& & & & & & & \\
(a)\0$\xi_8\h_8$ & (b)\0$\xi_3\h_7$ & $\sqrt{\frac{q^{-2}\lrpr{3}_q}{\lrpr{2}_q\lrpr{4}_q}}$ & $\sqrt{\frac{q^{-2}\lrpr{3}_q}{\lrpr{2}_q\lrpr{6}_q}}$ & \m$\frac{q}{\sqrt{\lrpr{2}_q\lrpr{4}_q}}$ & \m$\sqrt{\frac{q^4\lrpr{3}_q}{\lrpr{2}_q\lrpr{6}_q}}$ & $\frac{\sqrt{q^{-1}\lrpr{2}_q}}{\lrpr{4}_q}$ & \m$\frac{\lrpr{2}_q}{\lrpr{4}_q}\sqrt{\frac1{q^9\lrpr{6}_q}}$\\
& & & & & & & \\
(a)\0$\xi_9\h_7$ & (b)\0$\xi_2\h_8$ & $\sqrt{\frac{\lrpr{3}_q}{\lrpr{2}_q\lrpr{4}_q}}$ & \m$\sqrt{\frac{q^{-4}\lrpr{3}_q}{\lrpr{2}_q\lrpr{6}_q}}$ & \m$\frac{q^2}{\sqrt{\lrpr{2}_q\lrpr{4}_q}}$ & $\sqrt{\frac{q^2\lrpr{3}_q}{\lrpr{2}_q\lrpr{6}_q}}$ & $\frac{\sqrt{q^{-5}\lrpr{2}_q}}{\lrpr{4}_q}$ & $\frac{\lrpr{2}_q}{\lrpr{4}_q}\sqrt{\frac{1}{q^5\lrpr{6}_q}}$\\
& & & & & & & \\
(a)\0$\xi_8\h_7$ & (b)\0$\xi_6\h_4$ & 0 & 0 & 0 & 0 & $\frac{\lrpr{2}_q\sqrt{q\lrpr{2}_q}}{\lrpr{4}_q}$ & $\frac{\lrp{1-q^{-2}}\lrpr{2}_q}{\lrpr{4}_q\sqrt{q\lrpr{6}_q}}$\\
& & & & & & & \\
(a)\0$\xi_9\h_8$ & (b)\0$\xi_8\h_2$ & 0 & 0 & 0 & 0 & 0 & $\sqrt{\frac1{q^5\lrpr{6}_q}}$\\
& & & & & & & \\
(a)\0$\xi_9\h_5$ & (b)\0$\xi_9\h_1$ & 0 & 0 & 0 & 0 & 0 & $\sqrt{\frac1{q^3\lrpr{6}_q}}$\\
& & & & & & & \\
(a)\0$\xi_9\h_4$ & (b)\0$\xi_6\h_5$ & 0 & 0 & 0 & 0 & 0 & $\sqrt{\frac{\lrpr{3}_q}{\lrpr{6}_q}}$\\
\hline
\end{tabular}
\end{table*}

\begin{table*}
{TABLE III}: (Continued.)\vspace{0.5mm}\\    
\begin{tabular}{cccccccc}
\hline
& & (a)\hspace{0.35cm}$\psi^{\lrpy{51}}_{26}$ & $\psi^{\lrpy{42}}_{20}$ & $\psi^{\lrpy{42}}_{21}$ & $\psi^{\lrpy{3}}_{8}$ & $\psi^{\lrpy{21}}_{7}$ & (b)\hspace{0.1cm}$\psi^{\lrpy{21}}_{5}$ \\
\hline 
(a)\0$\xi_{10}\h_1$ & (b)\0$\xi_9\h_1$ & $q^3\sqrt{\frac{\lrpr{3}_q}{\lrpr{4}_q\lrpr{6}_q}}$ & $0$ & $\frac{q\lrpr{3}_q}{\sqrt{\lrpr{2}_q\lrpr{4}_q\lrpr{5}_q}}$ & $\sqrt{\frac{\lrpr{3}_q}{\lrpr{2}_q\lrpr{6}_q}}$ & $\sqrt{\frac{\lrpr{2}_q}{q^3\lrpr{5}_q}}$ & \m$\sqrt{\frac1{\lrpr{5}_q}}$\\
& & & & & & & \\
(a)\0$\xi_{9}\h_3$ & (b)\0$\xi_7\h_3$ & \m$\frac{q^{-1}}{\sqrt{\lrpr{4}_q\lrpr{6}_q}}$ & $\sqrt{\frac{\lrpr{2}_q}{\lrpr{3}_q\lrpr{4}_q}}$ & \m$\frac{q^{-1}\lrp{\lrpr{5}_q+q^{-1}\lrpr{2}_q}}{\sqrt{\lrpr{2}_q\lrpr{3}_q\lrpr{4}_q\lrpr{5}_q}}$\ & $\frac{q^2}{\sqrt{\lrpr{2}_q\lrpr{6}_q}}$ & $\sqrt{\frac{q\lrpr{2}_q}{\lrpr{3}_q\lrpr{5}_q}}$ & \m$\frac{q^{-2}}{\sqrt{\lrpr{5}_q}}$\\
& & & & & & & \\
(a)\0$\xi_5\h_8$ & (b)\0$\xi_5\h_6$ & $\frac{q^{-2}\lrpr{2}_q}{\sqrt{\lrpr{4}_q\lrpr{6}_q}}$ & \m$\sqrt{\frac{\lrpr{2}_q}{\lrpr{3}_q\lrpr{4}_q}}$ & \m$\sqrt{\frac{\lrpr{2}_q}{\lrpr{3}_q\lrpr{4}_q\lrpr{5}_q}}$ & \m$q\sqrt{\frac{\lrpr{2}_q}{\lrpr{6}_q}}$ & $\sqrt{\frac{q^5\lrpr{2}_q}{\lrpr{3}_q\lrpr{5}_q}}$ & $\frac{-q^{2}}{\sqrt{\lrpr{5}_q}}$\\
& & & & & & & \\
(a)\0$\xi_6\h_7$ & (b)\0$\xi_8\h_2$ & \m$\sqrt{\frac{\lrpr{2}_q}{q^7\lrpr{4}_q\lrpr{6}_q}}$ & \m$\frac{q^{3/2}\lrpr{2}_q}{\sqrt{\lrpr{3}_q\lrpr{4}_q}}$ & $\sqrt{\frac{1}{q^3\lrpr{3}_q\lrpr{4}_q\lrpr{5}_q}}$ & $\sqrt{\frac{1}{q\lrpr{6}_q}}$ & \m$\frac{q}{\sqrt{\lrpr{3}_q\lrpr{5}_q}}$ & $\frac{q^{-1}}{\sqrt{\lrpr{5}_q}}$\\
& & & & & & & \\
(a)\0$\xi_8\h_5$ & (b)\0$\xi_6\h_4$ & $\frac{\lrpr{3}_q\sqrt{q\lrpr{3}_q}}{\sqrt{\lrpr{2}_q\lrpr{4}_q\lrpr{6}_q}}$ & 0 & \m$\frac{\lrpr{3}_q}{\lrpr{2}_q}\sqrt{\frac{q^5}{\lrpr{4}_q\lrpr{5}_q}}$ & $\frac{\lrp{\lrpr{2}_q-q^3}\sqrt{\lrpr{3}_q}}{\lrpr{2}_q\sqrt{q^3\lrpr{6}_q}}$ & \m$\sqrt{\frac1{\lrpr{5}_q}}$ & $\frac{q}{\sqrt{\lrpr{5}_q}}$\\
& & & & & & & \\
(a)\0$\xi_8\h_4$ & (b)\0$\xi_2\h_8$ & $\sqrt{\frac{q}{\lrpr{2}_q\lrpr{4}_q\lrpr{6}_q}}$ & $\frac{\lrpr{2}_q}{\sqrt{q^3\lrpr{3}_q\lrpr{4}_q}}$ & $\frac{\lrpr{2}_q+q\lrpr{5}_q}{\lrpr{2}_q\sqrt{q\lrpr{3}_q\lrpr{4}_q\lrpr{5}_q}}$ & $\frac{-1}{\lrpr{2}_q}\sqrt{\frac{q^7}{\lrpr{6}_q}}$ & \m$\frac{q^2}{\sqrt{\lrpr{3}_q\lrpr{5}_q}}$ & 0\\
\hline
 & & (a)\hspace{0.5cm}$\psi^{\lrpy{42}}_{14}$ & $\psi^{\lrpy{3}}_{6}$ & $\psi^{\lrpy{21}}_{4}$ & (b)\hspace{0.5cm}$\psi^{\lrpy{42}}_{14}$ & $\psi^{\lrpy{3}}_{6}$ & $\psi^{\lrpy{21}}_{4}$\\
\hline
(a)\0$\xi_{8}\h_2$ & (b)\0$\xi_9\h_1$ & $\sqrt{\frac1{q\lrpr{2}_q\lrpr{5}_q}}$ & $\sqrt{\frac1{q^3\lrpr{6}_q}}$ & $\frac{q^{-3}}{\sqrt{\lrpr{3}_q\lrpr{5}_q}}$ & $\sqrt{\frac{q}{\lrpr{2}_q\lrpr{5}_q}}$ & $\sqrt{\frac{1}{q\lrpr{6}_q}}$ & $\frac{q^{-2}}{\sqrt{\lrpr{3}_q\lrpr{5}_q}}$\\
& & & & & & & \\
(a)\0$\xi_{2}\h_8$ & (b)\0$\xi_6\h_5$ & \m$\frac1{\lrpr{4}_q}\sqrt{\frac{\lrpr{2}_q}{q\lrpr{5}_q}}$ & \m$\sqrt{\frac{q}{\lrpr{6}_q}}$ & $\frac{q^{2}\lrpr{2}_q}{\sqrt{\lrpr{3}_q\lrpr{5}_q}}$ & \m$\sqrt{\frac{q^3\lrpr{3}_q}{\lrpr{2}_q\lrpr{5}_q}}$ & $\frac{q^{-5/2}-q^{-3/2}}{\lrpr{2}_q\sqrt{\lrpr{6}_q\lrp{\lrpr{3}_q}^{-1}}}$ & $\frac{-q^{-1}}{\sqrt{\lrpr{5}_q}}$\\
& & & & & & & \\
(a)\0$\xi_{6}\h_4$ & (b)\0$\xi_3\h_7$ & $\frac{\lrp{1+q^4-q^{-4}-q^2}\sqrt{\lrpr{2}_q}}{\lrpr{4}_q\sqrt{q\lrpr{5}_q}}$ & $\frac{q^{1/2}-q^{5/2}}{\sqrt{\lrpr{6}_q}}$ & $\frac{q^{-1}-q}{\sqrt{\lrpr{3}_q\lrpr{5}_q}}$ & $\frac{\sqrt{q^{-5}\lrpr{2}_q}}{\lrpr{4}_q\sqrt{\lrpr{5}_q}}$ & $\frac{\lrpr{3}_q+\lrpr{2}_q-q^5}{\lrp{\lrpr{2}_q}^2\sqrt{q^5\lrpr{6}_q}}$ & $\frac{-\lrpr{2}_q}{\sqrt{\lrpr{3}_q\lrpr{5}_q}}$\\
& & & & & & & \\
(a)\0$\xi_{5}\h_6$ & (b)\0$\xi_7\h_3$ & $\frac{\lrp{\lrpr{2}_q+q^5}\sqrt{\lrpr{2}_q}}{\lrpr{4}_q\sqrt{q^5\lrpr{5}_q}}$ & \m$\sqrt{\frac{q^3}{\lrpr{6}_q}}$ & $\frac{-1}{\sqrt{\lrpr{3}_q\lrpr{5}_q}}$ & $\frac{-\lrp{q^{-4}+q^{-2}+q^2}}{\lrpr{4}_q\sqrt{\lrpr{5}_q\lrp{q\lrpr{2}_q}^{-1}}}$ & $\sqrt{\frac{q^3}{\lrpr{6}_q}}$ & $\sqrt{\frac1{\lrpr{3}_q\lrpr{5}_q}}$\\
\hline
\end{tabular}
\end{table*}
\begin{table*}
{TABLE III}: (Continued.)\\\vspace{1mm} 
\begin{tabular}{cccccc}
\hline
& $\psi^{\lrpy{51}}_{28}$ & $\psi^{\lrpy{42}}_{22}$ & $\psi^{\lrpy{42}}_{23}$ & $\psi^{\lrpy{3}}_{9}$ & $\psi^{\lrpy{21}}_{8}$\\
\hline
$\xi_{10}\h_2$ & $q^3\sqrt{\frac{\lrpr{3}_q}{\lrpr{4}_q\lrpr{6}_q}}$ & $0$ & $\frac{q\lrpr{3}_q}{\sqrt{\lrpr{2}_q\lrpr{4}_q\lrpr{5}_q}}$ & $\sqrt{\frac{\lrpr{3}_q}{\lrpr{2}_q\lrpr{6}_q}}$ & $\sqrt{\frac{\lrpr{2}_q}{q^3\lrpr{5}_q}}$\\
& & & & & \\
$\xi_{8}\h_6$ & $\frac{1}{\sqrt{\lrpr{4}_q\lrpr{6}_q}}$ & $q^{-2}\sqrt{\frac{\lrpr{2}_q}{\lrpr{3}_q\lrpr{4}_q}}$ & \m$\frac{q^{-5/2}\lrp{\lrpr{5}_q+q\lrpr{2}_q}}{\sqrt{\lrpr{2}_q\lrpr{3}_q\lrpr{4}_q\lrpr{5}_q}}$ & \m$\frac{q^3}{\sqrt{\lrpr{2}_q\lrpr{6}_q}}$ & \m$\sqrt{\frac{q^3\lrpr{2}_q}{\lrpr{3}_q\lrpr{5}_q}}$\\
& & & & & \\
$\xi_7\h_7$ & \m$\frac{q^{-3}\lrpr{2}_q}{\sqrt{\lrpr{4}_q\lrpr{6}_q}}$ & \m$q^2\sqrt{\frac{\lrpr{2}_q}{\lrpr{3}_q\lrpr{4}_q}}$ & \m$q^{-1}\sqrt{\frac{\lrpr{2}_q}{\lrpr{3}_q\lrpr{4}_q\lrpr{5}_q}}$ & $\sqrt{\frac{\lrpr{2}_q}{\lrpr{6}_q}}$ & \m$\sqrt{\frac{q^3\lrpr{2}_q}{\lrpr{3}_q\lrpr{5}_q}}$\\
& & & & & \\
$\xi_6\h_8$ & $\sqrt{\frac{\lrpr{2}_q}{q^3\lrpr{4}_q\lrpr{6}_q}}$ & \m$\frac{q^{1/2}\lrpr{2}_q}{\sqrt{\lrpr{3}_q\lrpr{4}_q}}$ & $\sqrt{\frac{q}{\lrpr{3}_q\lrpr{4}_q\lrpr{5}_q}}$ & \m$\sqrt{\frac{q^3}{\lrpr{6}_q}}$ & $\frac{q^3}{\sqrt{\lrpr{3}_q\lrpr{5}_q}}$\\
& & & & & \\
$\xi_9\h_5$ & $\frac{\lrpr{3}_q\sqrt{q\lrpr{3}_q}}{\sqrt{\lrpr{2}_q\lrpr{4}_q\lrpr{6}_q}}$ & 0 & \m$\frac{\lrpr{3}_q}{\lrpr{2}_q}\sqrt{\frac{q^5}{\lrpr{4}_q\lrpr{5}_q}}$ & $\frac{\lrp{\lrpr{2}_q-q^3}\sqrt{\lrpr{3}_q}}{\lrpr{2}_q\sqrt{q^3\lrpr{6}_q}}$ & $\sqrt{\frac1{\lrpr{5}_q}}$\\
& & & & & \\
$\xi_9\h_4$ & \m$\sqrt{\frac{1}{q^3\lrpr{2}_q\lrpr{4}_q\lrpr{6}_q}}$ & $\frac{\lrpr{2}_q}{\sqrt{q\lrpr{3}_q\lrpr{4}_q}}$ & \m$\frac{\lrpr{2}_q+q\lrpr{5}_q}{\lrpr{2}_q\sqrt{q^5\lrpr{3}_q\lrpr{4}_q\lrpr{5}_q}}$ & $\frac{1}{\lrpr{2}_q}\sqrt{\frac{q^3}{\lrpr{6}_q}}$ & $\frac{1}{\sqrt{\lrpr{3}_q\lrpr{5}_q}}$\\
\hline
\end{tabular}
\end{table*}

\begin{table*}
{TABLE III}: (Continued.)\vspace{1mm}    
\begin{tabular}{ccccccc}
\hline
 &  \hspace{0cm} &(a)\hspace{0.5cm}$\psi^{\lrpy{51}}_{8}$\hspace{0.3cm} & \hspace{0.6cm}$\psi^{\lrpy{51}}_{7}$\hspace{0.4cm} &\hspace{0.7cm}$\psi^{\lrpy{42}}_{4}$\hspace{0.6cm} & \hspace{0.7cm}$\psi^{\lrpy{3}}_{1}$\hspace{0.6cm} &\hspace{0.3cm}(b)\hspace{0.2cm}$\psi^{\lrpy{42}}_{13}$ \\
\hline
(a)\0$\xi_{1}\h_4$ & (b)\0$\xi_8\h_2$ & $\frac{\lrpr{3}_q}{\sqrt{q^3\lrpr{2}_q\lrpr{4}_q\lrpr{5}_q\lrpr{6}_q}}$ & $\sqrt{\frac{\lrpr{2}_q}{q^3\lrpr{5}_q}}$ & $\frac1{\lrpr{2}_q}\sqrt{\frac{q\lrpr{3}_q}{\lrpr{4}_q}}$ & \m$\frac{\lrpr{3}_q}{\lrpr{2}_q}\sqrt{\frac{q^3}{\lrpr{6}_q}}$ & \m$\frac{\lrpr{3}_q}{\lrpr{4}_q}\sqrt{\frac{q^3\lrpr{2}_q}{\lrpr{3}_q}}$\\
& & & & & & \\
(a)\0$\xi_{1}\h_5$ & (b)\0$\xi_5\h_6$ & $\sqrt{\frac{\lrpr{3}_q\lrpr{5}_q}{q^3\lrpr{2}_q\lrpr{4}_q\lrpr{6}_q}}$ & 0 & \m$\frac{\lrpr{3}_q}{\lrpr{2}_q}\sqrt{\frac{q}{\lrpr{4}_q}}$ & \m$\frac1{\lrpr{2}_q}\sqrt{\frac{q^3\lrpr{3}_q}{\lrpr{6}_q}}$ & $\frac1{\lrpr{4}_q}\sqrt{\frac{\lrpr{2}_q}{q^7\lrpr{3}_q}}$\\
& & & & & & \\
(a)\0$\xi_2\h_3$ & (b)\0$\xi_7\h_3$ & \m$q^{-4}\sqrt{\frac{\lrpr{3}_q}{\lrpr{4}_q\lrpr{5}_q\lrpr{6}_q}}$ & $q\sqrt{\frac{\lrpr{3}_q}{\lrpr{5}_q}}$ & \m$q^{-2}\sqrt{\frac{1}{\lrpr{2}_q\lrpr{4}_q}}$ & $q^{-1}\sqrt{\frac{\lrpr{3}_q}{\lrpr{2}_q\lrpr{6}_q}}$ & $\frac1{\lrpr{4}_q}\sqrt{\frac{q\lrpr{2}_q}{\lrpr{3}_q}}$\\
& & & & & & \\
(a)\0$\xi_5\h_1$ & (b)\0$\xi_2\h_8$ & $q\sqrt{\frac{\lrpr{3}_q\lrpr{5}_q}{\lrpr{4}_q\lrpr{6}_q}}$ & 0 & $q^{-1}\sqrt{\frac{1}{\lrpr{2}_q\lrpr{4}_q}}$ & $q^{-2}\sqrt{\frac{\lrpr{3}_q}{\lrpr{2}_q\lrpr{6}_q}}$ & \m$\frac{\lrpr{3}_q}{\lrpr{4}_q}\sqrt{\frac{\lrpr{2}_q}{q\lrpr{3}_q}}$\\
& & & & & & \\
(a)\0$\xi_1\h_3$ & (b)\0$\xi_6\h_4$ & 0 & 0 & 0 & 0 & $\frac{\lrpr{2}_q}{\lrpr{4}_q}\sqrt{\frac{\lrpr{2}_q}{q^3\lrpr{3}_q}}$\\
\hline
& & (a)\hspace{0.2cm}$\psi^{\lrpy{51}}_{14}$ & \hspace{0.2cm}$\psi^{\lrpy{51}}_{13}$\hspace{0cm} &\hspace{0.3cm}$\psi^{\lrpy{42}}_{9}$ & \hspace{0.5cm}$\psi^{\lrpy{3}}_{4}$\hspace{0.4cm} &\hspace{0cm}(b)\hspace{0.2cm}$\psi^{\lrpy{42}}_{15}$ \\
\hline
(a)\0$\xi_{4}\h_4$ & (b)\0$\xi_8\h_2$ & $\frac{\lrpr{3}_q}{\sqrt{q^7\lrpr{2}_q\lrpr{4}_q\lrpr{5}_q\lrpr{6}_q}}$ & $\sqrt{\frac{q^3\lrpr{2}_q}{\lrpr{5}_q}}$ & \m$\frac1{\lrpr{2}_q}\sqrt{\frac{\lrpr{3}_q}{q^3\lrpr{4}_q}}$ & $\frac{\lrpr{3}_q}{\lrpr{2}_q}\sqrt{\frac{1}{q\lrpr{6}_q}}$ & $\sqrt{\frac{q^3\lrpr{3}_q}{\lrpr{2}_q\lrpr{5}_q}}$\\
& & & & & & \\
(a)\0$\xi_{4}\h_5$ & (b)\0$\xi_5\h_6$ & $\sqrt{\frac{\lrpr{3}_q\lrpr{5}_q}{q^3\lrpr{2}_q\lrpr{4}_q\lrpr{6}_q}}$ & 0 & \m$\frac{\lrpr{3}_q}{\lrpr{2}_q}\sqrt{\frac{q}{\lrpr{4}_q}}$ & \m$\frac1{\lrpr{2}_q}\sqrt{\frac{q^3\lrpr{3}_q}{\lrpr{6}_q}}$ & $\sqrt{\frac{\lrpr{2}_q}{q\lrpr{3}_q\lrpr{5}_q}}$\\
& & & & & & \\
(a)\0$\xi_3\h_6$ & (b)\0$\xi_7\h_3$ & $q^{-1}\sqrt{\frac{\lrpr{3}_q}{\lrpr{4}_q\lrpr{5}_q\lrpr{6}_q}}$ & $q^{-1}\sqrt{\frac{\lrpr{3}_q}{\lrpr{5}_q}}$ & $q\sqrt{\frac{1}{\lrpr{2}_q\lrpr{4}_q}}$ & \m$q^{2}\sqrt{\frac{\lrpr{3}_q}{\lrpr{2}_q\lrpr{6}_q}}$ & $\sqrt{\frac{\lrpr{2}_q}{q^5\lrpr{3}_q\lrpr{5}_q}}$\\
& & & & & & \\
(a)\0$\xi_7\h_2$ & (b)\0$\xi_9\h_1$ & $q\sqrt{\frac{\lrpr{3}_q\lrpr{5}_q}{\lrpr{4}_q\lrpr{6}_q}}$ & 0 & $q^{-1}\sqrt{\frac{1}{\lrpr{2}_q\lrpr{4}_q}}$ & $q^{-2}\sqrt{\frac{\lrpr{3}_q}{\lrpr{2}_q\lrpr{6}_q}}$ & \m$\sqrt{\frac{q\lrpr{3}_q}{\lrpr{2}_q\lrpr{5}_q}}$\\
& & & & & & \\
(a)\0$\xi_1\h_3$ & (b)\0$\xi_6\h_4$ & 0 & 0 & 0 & 0 & \m$\sqrt{\frac{\lrpr{2}_q}{q^3\lrpr{3}_q\lrpr{5}_q}}$\\
\hline
& $\psi^{\lrpy{51}}_{18}$ & $\psi^{\lrpy{51}}_{17}$ & $\psi^{\lrpy{42}}_{11}$ & $\psi^{\lrpy{42}}_{12}$ & $\psi^{\lrpy{3}}_{5}$ & $\psi^{\lrpy{21}}_{3}$ \\
\hline
$\xi_{5}\h_5$ & $\sqrt{\frac{\lrpr{3}_q}{q\lrpr{6}_q}}$ & 0 & 0 & \m$\sqrt{\frac{q^3\lrpr{3}_q}{\lrpr{2}_q\lrpr{5}_q}}$ & $\frac{q^{-5/2}-q^{-3/2}}{\lrpr{2}_q}\sqrt{\frac{\lrpr{3}_q}{\lrpr{6}_q}}$ & \m$\frac{q^{-1}}{\sqrt{\lrpr{5}_q}}$\\
& & & & & & \\
$\xi_{1}\h_8$ & $\frac{q^{-3}}{\lrpr{4}_q}\sqrt{\frac{\lrpr{2}_q\lrpr{3}_q}{\lrpr{6}_q}}$ & $\frac{q^{-3}}{\lrpr{4}_q}$ & \m$\frac{q^{-1}\sqrt{\lrpr{3}_q}}{\lrpr{4}_q}$ & \m$\frac{q^{-1}}{\lrpr{4}_q}\sqrt{\frac{\lrpr{3}_q}{\lrpr{5}_q}}$ & \m$\sqrt{\frac{\lrpr{3}_q}{\lrpr{2}_q\lrpr{6}_q}}$ & $\sqrt{\frac{q^3\lrpr{2}_q}{\lrpr{5}_q}}$\\
& & & & & & \\
$\xi_8\h_1$ & $q^{-2}\sqrt{\frac{\lrpr{2}_q}{\lrpr{6}_q}}$ & 0 & 0 & $\sqrt{\frac1{\lrpr{5}_q}}$ & $q^{-1}\sqrt{\frac{\lrpr{2}_q}{\lrpr{6}_q}}$ & $\sqrt{\frac{\lrpr{2}_q}{q^5\lrpr{3}_q\lrpr{5}_q}}$\\
& & & & & & \\
$\xi_2\h_7$ & \m$\frac{q^{-5}}{\lrpr{4}_q}\sqrt{\frac{\lrpr{2}_q}{\lrpr{6}_q}}$ & $\frac{\sqrt{\lrpr{3}_q}}{q\lrpr{4}_q}$ & \m$\frac{q\lrpr{3}_q}{\lrpr{4}_q}$ & $\frac{q^{-3}}{\lrpr{4}_q}\sqrt{\frac1{\lrpr{5}_q}}$ & $\frac{\lrpr{3}_q+\lrp{1-q^{3/2}}\lrpr{2}_q-q^5}{\lrpr{2}_q\sqrt{q^3\lrpr{2}_q\lrpr{6}_q}}$ & \m$\sqrt{\frac{\lrpr{2}_q}{q\lrpr{3}_q\lrpr{5}_q}}$\\
& & & & & & \\
$\xi_5\h_4$ & $\frac{\lrpr{2}_q}{\lrpr{4}_q}\sqrt{\frac1{q^5\lrpr{6}_q}}$ & $\frac{\sqrt{\lrpr{2}_q\lrpr{3}_q}}{q^{1/2}\lrpr{4}_q}$ & $\frac{q^{-5/2}\sqrt{\lrpr{2}_q}}{\lrpr{4}_q}$ & $\frac{\lrpr{2}_q+q^5}{\lrpr{4}_q}\sqrt{\frac{\lrpr{2}_q}{q^3\lrpr{5}_q}}$ & \m$\sqrt{\frac{q^5}{\lrpr{6}_q}}$ & \m$\frac{q}{\sqrt{\lrpr{3}_q\lrpr{5}_q}}$\\
& & & & & & \\
$\xi_6\h_3$ & \m$\frac{\lrpr{2}_q}{\lrpr{4}_q}\sqrt{\frac1{q\lrpr{6}_q}}$ & $\frac{\sqrt{q^3\lrpr{2}_q\lrpr{3}_q}}{\lrpr{4}_q}$ & $\frac{q^{-1/2}\sqrt{\lrpr{2}_q}}{\lrpr{4}_q}$ & \m$\frac{q^{-4}+q^{-2}+q^2}{\lrpr{4}_q}\sqrt{\frac{\lrpr{2}_q}{q\lrpr{5}_q}}$ & $\sqrt{\frac{q}{\lrpr{6}_q}}$ & $\frac{q^{-1}}{\sqrt{\lrpr{3}_q\lrpr{5}_q}}$\\
\hline
\end{tabular}
\end{table*}

\begin{table*}
{TABLE III}: (Continued.)\vspace{0.5mm}\\    
\begin{tabular}{ccccccc}
\hline
& $\psi^{\lrpy{51}}_{9}$ & $\psi^{\lrpy{51}}_{10}$ & $\psi^{\lrpy{42}}_{5}$ & $\psi^{\lrpy{42}}_{6}$ & $\psi^{\lrpy{3}}_{2}$ & $\psi^{\lrpy{21}}_{1}$ \\
\hline
$\xi_{3}\h_3$ & $q^2\sqrt{\frac{\lrpr{2}_q\lrpr{3}_q}{\lrpr{4}_q\lrpr{5}_q}}$ & \m$\frac{q^{-3}\lrpr{2}_q}{\sqrt{\lrpr{4}_q\lrpr{5}_q\lrpr{6}_q}}$ & \m$\frac{q^{-1}\sqrt{\lrpr{2}_q}}{\sqrt{\lrpr{3}_q\lrpr{4}_q}}$ & $\frac{q^{-4}\sqrt{\lrpr{2}_q}}{\sqrt{\lrpr{3}_q\lrpr{4}_q\lrpr{5}_q}}$ & $\sqrt{\frac{\lrpr{2}_q}{\lrpr{6}_q}}$ & \m$\sqrt{\frac{\lrpr{2}_q}{q^3\lrpr{3}_q\lrpr{5}_q}}$\\
& & & & & & \\
$\xi_{6}\h_1$ & 0 & $\sqrt{\frac{q^3\lrpr{2}_q\lrpr{5}_q}{\lrpr{4}_q\lrpr{6}_q}}$ & $\sqrt{\frac{1}{q\lrpr{3}_q\lrpr{4}_q}}$ & \m$\frac{\lrpr{4}_q+q^{-1}-1}{\sqrt{\lrpr{3}_q\lrpr{4}_q\lrpr{5}_q}}$ & $\sqrt{\frac1{q^3\lrpr{6}_q}}$ & \m$\frac{q^{-3}}{\sqrt{\lrpr{3}_q\lrpr{5}_q}}$\\
& & & & & & \\
$\xi_{2}\h_4$ & $\frac{\lrpr{2}_q\sqrt{\lrpr{3}_q}}{\sqrt{q\lrpr{4}_q\lrpr{5}_q}}$ & $\frac{q^{-1/2}\lrp{\lrpr{2}_q-q^{-5}}}{\sqrt{\lrpr{2}_q\lrpr{4}_q\lrpr{5}_q\lrpr{6}_q}}$ & $\frac{q^{3/2}\lrp{\lrpr{2}_q-q^{-4}}}{\lrpr{2}_q\sqrt{\lrpr{3}_q\lrpr{4}_q}}$ & \m$\frac{\lrpr{2}_q}{\sqrt{q^5\lrpr{3}_q\lrpr{4}_q\lrpr{5}_q}}$ & $\frac{q^{5/2}\lrp{q^{-5}-\lrpr{2}_q}}{\lrpr{2}_q\sqrt{\lrpr{6}_q}}$ & $\frac{\lrpr{2}_q}{\sqrt{\lrpr{3}_q\lrpr{5}_q}}$\\
& & & & & & \\
$\xi_{5}\h_2$ & 0 & $\sqrt{\frac{\lrpr{5}_q}{\lrpr{4}_q\lrpr{6}_q}}$ & $\frac{q^{-2}}{\sqrt{\lrpr{2}_q\lrpr{3}_q\lrpr{4}_q}}$ & $\frac{q^3\lrp{1+q^3\lrpr{3}_q}\sqrt{\lrpr{2}_q}}{\sqrt{\lrpr{3}_q\lrpr{4}_q\lrpr{5}_q}}$ & $\frac{q^{-3}}{\sqrt{\lrpr{2}_q\lrpr{6}_q}}$ & $\sqrt{\frac{\lrpr{2}_q}{q^3\lrpr{3}_q\lrpr{5}_q}}$\\& & & & & & \\
$\xi_{1}\h_6$ & $\frac{q^{-3}\sqrt{\lrpr{2}_q}}{\sqrt{\lrpr{4}_q\lrpr{5}_q}}$ & $\frac{q^{-3}\sqrt{\lrpr{3}_q}}{\sqrt{\lrpr{4}_q\lrpr{5}_q\lrpr{6}_q}}$ & \hspace{0.5cm}$\frac{q^{-1}}{\sqrt{\lrpr{2}_q\lrpr{4}_q}}$ & $\frac{q^{-1}\sqrt{\lrpr{2}_q}}{\sqrt{\lrpr{4}_q\lrpr{5}_q}}$ & \m$\sqrt{\frac{\lrpr{3}_q}{\lrpr{2}_q\lrpr{6}_q}}$ & \m$\sqrt{\frac{q^3\lrpr{2}_q}{\lrpr{5}_q}}$\\
& & & & & & \\
$\xi_{2}\h_5$ & 0 & $\sqrt{\frac{\lrpr{3}_q\lrpr{5}_q}{q^3\lrpr{2}_q\lrpr{4}_q\lrpr{6}_q}}$ &\hspace{0.5cm}\m$\frac{\lrpr{3}_q}{\lrpr{2}_q}\sqrt{\frac{q}{\lrpr{4}_q}}$ & 0 & \m$\frac1{\lrpr{2}_q}\sqrt{\frac{q^3\lrpr{3}_q}{\lrpr{6}_q}}$ & 0\\
\hline
& $\psi^{\lrpy{51}}_{22}$ & $\psi^{\lrpy{51}}_{21}$ & $\psi^{\lrpy{42}}_{16}$ & $\psi^{\lrpy{42}}_{17}$ & $\psi^{\lrpy{3}}_{7}$ & $\psi^{\lrpy{21}}_{6}$ \\
\hline
$\xi_{7}\h_5$ & $\sqrt{\frac{\lrpr{3}_q}{q\lrpr{6}_q}}$ & 0 & 0 & \m$\sqrt{\frac{q^3\lrpr{3}_q}{\lrpr{2}_q\lrpr{5}_q}}$ & $\frac{q^{-5/2}-q^{-3/2}}{\lrpr{2}_q}\sqrt{\frac{\lrpr{3}_q}{\lrpr{6}_q}}$ & \m$\frac{q^{-1}}{\sqrt{\lrpr{5}_q}}$\\
& & & & & & \\
$\xi_{4}\h_7$ & $\frac{-q^{-4}}{\lrpr{4}_q}\sqrt{\frac{\lrpr{2}_q\lrpr{3}_q}{\lrpr{6}_q}}$ & $\frac{1}{\lrpr{4}_q}$ & \m$\frac{q^{2}\sqrt{\lrpr{3}_q}}{\lrpr{4}_q}$ & $\frac{q^{-2}}{\lrpr{4}_q}\sqrt{\frac{\lrpr{3}_q}{\lrpr{5}_q}}$ & $\frac{1+q^{-1}}{q^2\lrpr{2}_q}\sqrt{\frac{\lrpr{3}_q}{\lrpr{2}_q\lrpr{6}_q}}$ & \m$\sqrt{\frac{q\lrpr{2}_q}{\lrpr{5}_q}}$\\
& & & & & & \\
$\xi_9\h_2$ & $q^{-2}\sqrt{\frac{\lrpr{2}_q}{\lrpr{6}_q}}$ & 0 & 0 & $\sqrt{\frac1{\lrpr{5}_q}}$ & $q^{-1}\sqrt{\frac{\lrpr{2}_q}{\lrpr{6}_q}}$ & $\sqrt{\frac{\lrpr{2}_q}{q^5\lrpr{3}_q\lrpr{5}_q}}$\\
& & & & & & \\
$\xi_3\h_8$ & $\frac{q^{-2}}{\lrpr{4}_q}\sqrt{\frac{\lrpr{2}_q}{\lrpr{6}_q}}$ & $\frac{\sqrt{\lrpr{3}_q}}{q^2\lrpr{4}_q}$ & \m$\frac{\lrpr{3}_q}{\lrpr{4}_q}$ & \m$\frac{1}{\lrpr{4}_q}\sqrt{\frac1{\lrpr{5}_q}}$ & \m$q\sqrt{\frac1{\lrpr{2}_q\lrpr{6}_q}}$ & $\sqrt{\frac{q^3\lrpr{2}_q}{\lrpr{3}_q\lrpr{5}_q}}$\\
& & & & & & \\
$\xi_6\h_6$ & $\frac{\lrpr{2}_q}{\lrpr{4}_q}\sqrt{\frac1{q\lrpr{6}_q}}$ & $\frac{\sqrt{\lrpr{2}_q\lrpr{3}_q}}{q^{5/2}\lrpr{4}_q}$ & $\frac{q^{-5/2}\sqrt{\lrpr{2}_q}}{\lrpr{4}_q}$ & $\frac{\lrpr{5}_q+q^{-2}}{\lrpr{2}_q\lrpr{4}_q}\sqrt{\frac{q\lrpr{2}_q}{\lrpr{5}_q}}$ & \m$\sqrt{\frac{q^5}{\lrpr{6}_q}}$ & \m$\frac{q}{\sqrt{\lrpr{3}_q\lrpr{5}_q}}$\\
& & & & & & \\
$\xi_7\h_4$ & \m$\frac{\lrpr{2}_q}{\lrpr{4}_q}\sqrt{\frac1{q^5\lrpr{6}_q}}$ & $\frac{\sqrt{q^3\lrpr{2}_q\lrpr{3}_q}}{\lrpr{4}_q}$ & $\frac{q^{-1/2}\sqrt{\lrpr{2}_q}}{\lrpr{4}_q}$ & \m$\frac{q^{-4}+q^{-2}+q^2}{\lrpr{4}_q}\sqrt{\frac{\lrpr{2}_q}{q\lrpr{5}_q}}$ & $\sqrt{\frac{q}{\lrpr{6}_q}}$ & $\frac{q^{-1}}{\sqrt{\lrpr{3}_q\lrpr{5}_q}}$\\
\hline
& $\psi^{\lrpy{51}}_{11}$ & $\psi^{\lrpy{51}}_{12}$ & $\psi^{\lrpy{42}}_{7}$ & $\psi^{\lrpy{42}}_{8}$ & $\psi^{\lrpy{3}}_{3}$ & $\psi^{\lrpy{21}}_{2}$ \\
\hline 
$\xi_{2}\h_6$ & $q^{-2}\sqrt{\frac{\lrpr{2}_q\lrpr{3}_q}{\lrpr{4}_q\lrpr{5}_q}}$ & $\frac{q^{-2}\lrpr{2}_q}{\sqrt{\lrpr{4}_q\lrpr{5}_q\lrpr{6}_q}}$ & $\sqrt{\frac{\lrpr{2}_q}{\lrpr{3}_q\lrpr{4}_q}}$ & $\sqrt{\frac{\lrpr{2}_q}{\lrpr{3}_q\lrpr{4}_q\lrpr{5}_q}}$ & \m$q\sqrt{\frac{\lrpr{2}_q}{\lrpr{6}_q}}$ & \m$\sqrt{\frac{q^5\lrpr{2}_q}{\lrpr{3}_q\lrpr{5}_q}}$\\
& & & & & & \\
$\xi_{6}\h_2$ & 0 & $\sqrt{\frac{q\lrpr{2}_q\lrpr{5}_q}{\lrpr{4}_q\lrpr{6}_q}}$ & $\sqrt{\frac{1}{q\lrpr{3}_q\lrpr{4}_q}}$ & $\frac{q^{-1/2}\lrp{1+q^3\lrpr{3}_q}}{\sqrt{\lrpr{3}_q\lrpr{4}_q\lrpr{5}_q}}$ & $\sqrt{\frac1{q^5\lrpr{6}_q}}$ & $\frac{q^{-1}}{\sqrt{\lrpr{3}_q\lrpr{5}_q}}$\\
& & & & & & \\
$\xi_{3}\h_4$ & $\frac{\lrpr{2}_q\sqrt{q\lrpr{3}_q}}{\sqrt{\lrpr{4}_q\lrpr{5}_q}}$ & $\frac{q^{1/2}\lrp{1-q^{-5}\lrpr{2}_q}}{\sqrt{\lrpr{2}_q\lrpr{4}_q\lrpr{5}_q\lrpr{6}_q}}$ & $\frac{q^{-5/2}\lrp{q^5-\lrpr{2}_q}}{\lrpr{2}_q\sqrt{\lrpr{3}_q\lrpr{4}_q}}$ & \m$\frac{\lrpr{2}_q}{\sqrt{q^3\lrpr{3}_q\lrpr{4}_q\lrpr{5}_q}}$ & $\frac{q^{-3/2}\lrp{\lrpr{2}_q-q^5}}{\lrpr{2}_q\sqrt{\lrpr{6}_q}}$ & $\frac{q\lrpr{2}_q}{\sqrt{\lrpr{3}_q\lrpr{5}_q}}$\\
\hline
\end{tabular}
\end{table*}

\begin{table*}
{TABLE III}: (Continued.)\\\vspace{1mm} 
\begin{tabular}{ccccccc}
\hline
$\xi_{7}\h_1$ & 0 & $q^2\sqrt{\frac{\lrpr{5}_q}{\lrpr{4}_q\lrpr{6}_q}}$ & $\frac{1}{\sqrt{\lrpr{2}_q\lrpr{3}_q\lrpr{4}_q}}$ & \m$\sqrt{\frac{\lrpr{2}_q\lrpr{4}_q}{\lrpr{3}_q\lrpr{5}_q}}$ & $\frac{q^{-1}}{\sqrt{\lrpr{2}_q\lrpr{6}_q}}$ & \m$\sqrt{\frac{\lrpr{2}_q}{q^5\lrpr{3}_q\lrpr{5}_q}}$\\
& & & & & & \\
$\xi_{4}\h_3$ & $\frac{q^{3}\sqrt{\lrpr{2}_q}}{\sqrt{\lrpr{4}_q\lrpr{5}_q}}$ & $\frac{-q^{-2}\sqrt{\lrpr{3}_q}}{\sqrt{\lrpr{4}_q\lrpr{5}_q\lrpr{6}_q}}$ & \m$\frac{1}{\sqrt{\lrpr{2}_q\lrpr{4}_q}}$ & $\frac{q^{-3}\sqrt{\lrpr{2}_q}}{\sqrt{\lrpr{4}_q\lrpr{5}_q}}$ & $q\sqrt{\frac{\lrpr{3}_q}{\lrpr{2}_q\lrpr{6}_q}}$ & \m$\sqrt{\frac{\lrpr{2}_q}{q\lrpr{5}_q}}$\\
& & & & & & \\
$\xi_{3}\h_5$ & 0 & $\sqrt{\frac{\lrpr{3}_q\lrpr{5}_q}{q^3\lrpr{2}_q\lrpr{4}_q\lrpr{6}_q}}$ & \m$\frac{\lrpr{3}_q}{\lrpr{2}_q}\sqrt{\frac{q}{\lrpr{4}_q}}$ & 0 & \m$\frac1{\lrpr{2}_q}\sqrt{\frac{q^3\lrpr{3}_q}{\lrpr{6}_q}}$ & 0\\
\hline
\end{tabular}
\end{table*}
\bibliography{apssamp}

\end{document}